\documentclass[a4paper,onecolumn,nofootinbib,superscriptaddress,prd]{revtex4-1}
\usepackage{graphicx}
\usepackage{amsmath}
\usepackage{amssymb}
\usepackage{xcolor}
\usepackage{bm}
\usepackage{aas_macros}
\usepackage{hyperref}

\newcommand{\vk}{\vec k}
\newcommand{\vq}{\vec q}
\newcommand{\vp}{\vec p}
\renewcommand{\vec}{\bm}
\newcommand{\be}{\begin{equation}}
\newcommand{\ee}{\end{equation}}
\newcommand{\hMpc}{h^{-1}\ \text{Mpc}}
\newcommand{\hMpcsq}{h^{-2}\ \text{Mpc}^2}
\newcommand{\ihMpc}{h\ \text{Mpc}^{-1}}

\newcommand{\cssq}{c_\text{s}^2}
\newcommand{\sigmadsq}{\sigma_\text{d}^2}

\newcommand{\bea}{\begin{eqnarray}}
\newcommand{\eea}{\end{eqnarray}}
\newcommand{\nn}{\nonumber}
\newcommand{\tr}{\mbox{tr}}

\renewcommand{\vec}[1]{{\bm #1}}

\begin{document}

\title{The two-loop bispectrum of large-scale structure}
\author{Tobias Baldauf}
\email{t.baldauf@tbaweb.de}
\affiliation{Department of Applied Mathematics and Theoretical Physics, University of Cambridge, Wilberforce Road, CB3 0WA}
\author{Mathias Garny}
\email{mathias.garny@tum.de}
\affiliation{Physik Department T31, Technische Universit\"at M\"unchen, James-Franck-Stra\ss{}e 1, D-85748 Garching,
Germany}
\author{Petter Taule}
\email{petter.taule@tum.de}
\affiliation{Physik Department T31, Technische Universit\"at M\"unchen, James-Franck-Stra\ss{}e 1, D-85748 Garching,
Germany}
\author{Theo Steele}
\email{ts715@damtp.cam.ac.uk}
\affiliation{Department of Applied Mathematics and Theoretical Physics, University of Cambridge, Wilberforce Road, CB3 0WA}


\hfill TUM-HEP-1360/21

\hfill \today

\begin{abstract}
The bispectrum is the leading non-Gaussian statistic in large-scale structure, carrying valuable information on cosmology that is complementary to the power spectrum. To access this information, we need to model the bispectrum in the weakly non-linear regime. In this work we present the first two-loop, i.e., next-to-next-to-leading order perturbative description of the bispectrum within an effective field theory (EFT) framework. Using an analytic expansion of the perturbative kernels up to $F_6$ we derive a renormalized bispectrum that is demonstrated to be independent of the UV cutoff. We show that the EFT parameters associated with the four independent second-order EFT operators known from the one-loop bispectrum are sufficient to absorb the UV sensitivity of the two-loop contributions in the double-hard region. In addition, we employ a simplified treatment of the single-hard region, introducing one extra EFT parameter at two-loop order. We compare our results to N-body simulations using the realization-based grid-PT method and find good agreement within the expected range, as well as consistent values for the EFT parameters. The two-loop terms start to become relevant at $k\approx 0.07\ihMpc$. The range of wavenumbers with percent-level agreement, independently of the shape, extends from $0.08\ihMpc$ to $0.15\ihMpc$ when going from one to two loops at $z=0$. In addition, we quantify the impact of using exact instead of Einstein-de-Sitter kernels for the one-loop bispectrum, and discuss in how far their impact can be absorbed into a shift of the EFT parameters.
\end{abstract}

\maketitle

\tableofcontents

\section{Introduction}

Large-scale structure (LSS) surveys will extend to increasingly larger scales as well as higher redshifts in the near future, thereby moving more and more into the weakly non-linear regime of structure formation. Perturbative methods augmented with models that capture uncertainties from the impact of strongly non-linear effects already play an important role for extracting information on cosmology  from survey data, and will become increasingly powerful with the ongoing observational progress and advance in theoretical understanding.  Present and near future surveys such as eBOSS \cite{Zhao:2015gua}, Euclid \cite{2011arXiv1110.3193L}, DES \cite{DES:2016jjg}, DESI \cite{Aghamousa:2016zmz}, HETDEX \cite{2021ApJ...911..108H}, HSC \cite{HSC:2018mrq}, KiDS \cite{2013ExA....35...25D}, the Vera C. Rubin Observatory (LSST) \cite{2019ApJ...873..111I}, 4MOST \cite{deJong:2012nj}, PFS \cite{2014PASJ...66R...1T}, eROSITA \cite{eROSITA:2012lfj}, SPHEREx \cite{Dore:2014cca}, VIPERS \cite{Pezzotta:2016gbo}, and the Nancy Grace Roman Space Telescope (WFIRST) \cite{2021MNRAS.tmp.1608E} are allowing for data comparisons that will provide unprecedented levels of precision in our knowledge of fundamental physical parameters. 

Within the framework of standard perturbation theory (SPT), the evolution of the density contrast $\delta$ is described as a pressureless perfect fluid governed by a continuity and Euler equation~\cite{Bernardeau:2001qr}. The equations can be viewed as a truncation of the coupled hierarchy of moments of the phase-space distribution function at the first order, i.e.\ following the evolution of the density and velocity fields. Within SPT, the contribution from the second moment, the velocity dispersion tensor, to the Euler equation is neglected. This is motivated by its initial smallness for cold dark matter. Nevertheless, it is well-known that non-linear evolution generates a significant velocity dispersion~\cite{Pueblas:2008uv}, which in turn back-reacts on the velocity and density fields. As a consequence, SPT becomes inaccurate on small scales, which then impacts the observable modes in the range of baryon acoustic oscillations (BAO) due to non-linear mode-coupling. This inaccuracy shows up in practice as a certain dependence of power- and bispectra predictions at higher order in perturbation theory on fluctuations with large wavenumber, known as \emph{UV sensitivity}. This UV sensitivity becomes increasingly dominant at higher orders in the perturbative expansion~\cite{Blas:2013aba,Konstandin:2019bay}, and signals the breakdown of the perfect pressureless fluid description.

Over the last decade an effective field theory (EFT) approach has been proposed to systematically parametrize the impact of small-scale, non-perturbative effects onto larger, perturbative scales~\cite{Baumann:2010tm,Carrasco:2012cv,Pajer:2013jj}.
The EFT description is obtained by coarse-graining the density and velocity field with a smoothing scale $\Lambda$, such that fluctuations on smaller scales are ``integrated out''. The evolution of the smoothed fields can be described by a modified Euler equation, that contains an effective stress tensor~\cite{Abolhasani:2015mra}. The latter is treated as a functional that depends on (gradients of) the smoothed fields as well as stochastic terms, in close analogy to the formalism developed to describe biased tracers~\cite{Kaiser:1984sw,Bardeen:1985tr,Fry:1992vr,Scherrer:1997hp,Matsubara:1999qq,Chan:2012jj,Desjacques:2016bnm}. The effective stress tensor has in general the form of a sum of ``operators'', being products of gradients of the smoothed gravitational potential and velocity fields that are compatible with Galilean symmetry~\cite{Abolhasani:2015mra}. Each operator is multiplied by an a priori unknown coefficient. These parameters of the EFT, which are also referred to as low-energy constants or Wilson coefficients, are treated as free parameters that need to be constrained in simulations or marginalized over when constraining physically relevant parameters. 

\begin{figure}
    \centering
     \includegraphics[width=0.6\textwidth]{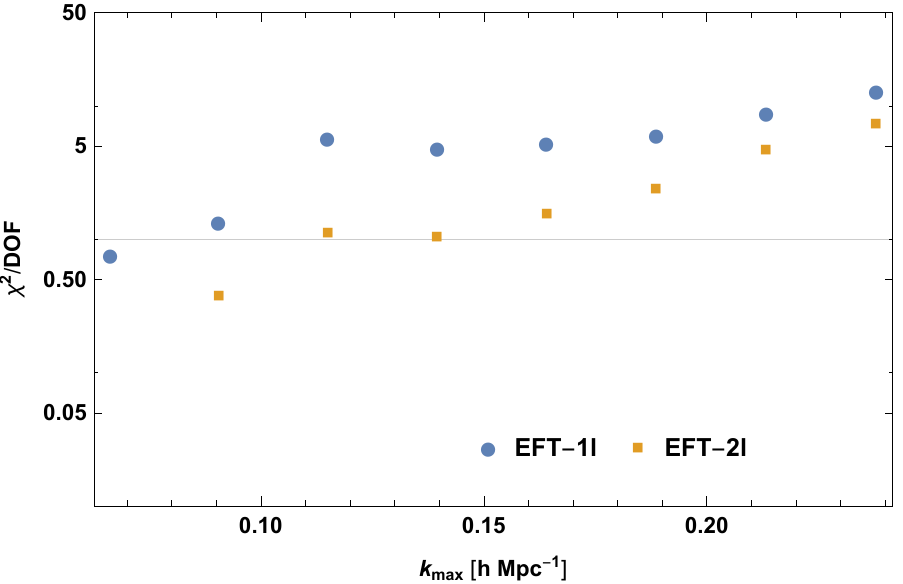}
   \caption{ \label{fig:equilat_summary} 
    $\chi^2$ per degree of freedom (as defined in~\eqref{chinnn}) obtained from a fit of the one-(two-)loop bispectrum to N-body results at $z=0$ involving four (five) EFT parameters. The fit encompasses a full set of configurations $B(k_1,k_2,k_3)$ with $k_i\leq k_\text{max}$, and uses the grid-PT method to alleviate cosmic variance on large scales. The range of scale for which $\chi^2/$DOF~$<1$ increases from $0.08 \ihMpc$ to $0.15\ihMpc $ when including two-loop (NNLO) contributions.}
\end{figure}

In general, since both large and small scales evolve on Hubble time-scales, the operators furnishing the EFT expansion of the effective stress tensor are non-local in time, and contain an integral over the past evolution of the system~\cite{Abolhasani:2015mra}. Alternatively, one may include a set of operators encompassing an arbitrary number of time-derivatives. Due to the absence of a separation of time-scales for the large and small scales, adding time derivatives does not in general lead to a suppression of the operator~\cite{Mirbabayi:2014zca}. Obviously, including an infinite set of operators, that all contribute with a comparable magnitude, would render the EFT expansion useless. Nevertheless, under certain conditions, it is possible to express the effective stress tensor in terms of a \emph{finite} number of local-in-time operators, when working to a certain order in perturbation theory \cite{Mirbabayi:2014zca,Abolhasani:2015mra,Desjacques:2016bnm}. This is strictly the case if the time-dependence of the hard modes that are being integrated out can be expressed purely in powers of the linear growth factor, such as occurs in SPT when adopting the Einstein-de-Sitter (EdS) approximation. In reality, apart from deviations from the EdS approximation, the UV modes are subject to complex non-linear processes, including also baryonic effects, leading to the appearance of additional time-scales. Nevertheless, in practice, choosing a subset of operators that are designed to correct for the UV sensitivity of SPT, it has been shown that perturbative predictions of the power spectrum can be brought into closer agreement with simulation measurements. 

Beyond the power spectrum, the bispectrum provides complementary information
and is instrumental in disentangling bias from the impact of fundamental parameters~\cite{Scoccimarro:1997st,Scoccimarro:2000sn,Sefusatti:2006pa,Gil-Marin:2016wya,Ruggeri:2017dda,Eggemeier:2018qae,Hahn:2019zob,Oddo:2019run,Eggemeier:2021cam,Oddo:2021iwq,Ivanov:2021kcd}.
Within an EFT framework the matter bispectrum has been studied in \cite{Baldauf:2014qfa,Angulo:2014tfa}.
Recently, precision tests of the one-loop bispectrum \cite{Steele:2020tak,Alkhanishvili:2021pvy} have pointed out that the range of validity of the one-loop EFT is much more restricted than expected by initial studies. This is in line with findings for the two-loop power spectrum \cite{Baldauf:2015aha} and expectations based on theoretical errors \cite{Baldauf:2016sjb}. In \cite{Steele:2020tak} it was also pointed out, that the common EdS approximation for the non-linear $F_2$ kernel leads to errors that exceed the size of the one-loop bispectrum on large scales. 
Finally, cubic interactions in the matter trispectrum \cite{Gualdi:2020eag} were studied in \cite{Bertolini:2015fya,Bertolini:2016bmt} based on the covariance matrix and more recently by \cite{Steele:2021} in explicit measurements of the trispectrum.

In this work we present the first study of the two-loop, i.e.\ next-to-next-to-leading order (NNLO), bispectrum including EFT corrections that account for the spurious UV sensitivity of SPT to this order in perturbation theory. In addition, we evaluate the effect of the exact $\Lambda$CDM non-linear $F_4$ kernel on the one-loop contribution to the bispectrum for the first time. We evaluate the perturbative predictions in two different ways: based on a direct computation using Monte Carlo integration with the algorithm outlined in \cite{Blas:2013bpa,Blas:2013aba,Floerchinger:2019eoj}, that accounts for the cancellation of infrared-enhanced contributions on the integrand level, as well as using a realization-based approach (known as grid-PT)~\cite{Roth:2011test,Taruya:2018jtk,Taruya:2020qoy}. The latter is used to compare to numerical N-body simulations, and calibrate the EFT corrections.

A summary of our results is shown in Fig.\,\ref{fig:equilat_summary}, where we compare EFT one- and two-loop bispectra with N-body simulations results at $z=0$. 
The EFT bispectra involve in total four (five) free parameters at one- and two-loop order, respectively. 
To obtain the EFT parameters we used a set of configurations $B(k_1,k_2,k_3)$, and we present the detailed setup and procedure below. Fig.\,\ref{fig:equilat_summary} shows the $\chi^2$ per degree of freedom when taking our full set of configurations into account, with each of the three wavenumbers below a maximal value $k_\text{max}$. We find that $\chi^2/$DOF is below unity for $k_\text{max}\lesssim 0.08\ihMpc$ for the EFT one-loop result, and for $k_\text{max}\lesssim 0.15\ihMpc$ for the EFT two-loop approximation that comprises the main result of this work.

This work is structured as follows. After reviewing the logic of the effective field theory approach in section~\ref{sec:eft} as well as its application to the one-loop bispectrum in section~\ref{sec:oneloop}, we present results for the one-loop bispectrum with exact $\Lambda$CDM non-linear kernels in section~\ref{sec:exactoneloop}. The two-loop bispectrum as well as our treatment of EFT corrections is discussed in section~\ref{sec:twoloop}, while a comparison to N-body data is presented in section~\ref{sec:numerics}. We conclude in section~\ref{sec:conclusions}.

\section{Review: Effective field theory setup}\label{sec:eft}

While the Universe is seeded by almost perfectly Gaussian initial conditions, the growth of structure eventually leads to non-linear mode coupling. Within the EFT treatment, the evolution of the coarse-grained density and velocity fields is described by a continuity and Euler equation, including a source term in the form of an effective stress tensor. In Fourier space the dynamical equations for the density contrast $\delta$ and velocity divergence $\theta=\partial_i \vec v^i$ are thus given by
\begin{align}
\delta'(\vk)+\theta(\vk)=&-\int d^3q\,\alpha(\vq,\vk-\vq)\theta(\vq)\delta(\vk-\vq)\,,\\
\theta'(\vk)+\mathcal{H}\theta(\vk) +\frac{3}{2}\Omega_\text{m}\mathcal{H}^2\delta(\vk)=&-\int d^3q\,\beta(\vq,\vk-\vq)\theta(\vq)\theta(\vk-\vq) - \tau_\theta(\vk)\,,
\end{align}
where $\alpha(\vq,\vk-\vq)=\vq\cdot\vk/q^2$ and $\beta(\vq,\vk-\vq)=\vq\cdot(\vk-\vq)k^2/(2q^2(\vk-\vq)^2)$ are the mode coupling functions, $\Omega_\text{m}$ is the time-dependent matter density parameter, ${\cal H}=aH$ is the conformal Hubble rate, and the prime denotes a derivative with respect to conformal time $\tau$.

The effective stress term 
\be\label{eq:tautheta}
  \tau_\theta = \partial_i\frac{1}{1+\delta}\partial_j\tau^{ij}\,,
\ee
where $\tau^{ij}$ is the effective stress tensor,
provides EFT corrections to capture the short wavelength deviations from the pressureless perfect fluid assumed in the SPT treatment.
The factor of $1/(1+\delta)$ in the EFT source term $\tau_\theta$ defined in \eqref{eq:tautheta} ensures that the density contrast scales with total wavenumber squared in the large-scale limit, as required by momentum conservation~\cite{Abolhasani:2015mra}. This can also be seen directly by taking a time-derivative of the continuity equation, and using the Euler equation, giving
\be 
  \delta''+{\cal H}\delta'-\frac32\Omega_\text{m}{\cal H}^2\delta = \partial_i\partial_j[(1+\delta)\vec v^i\vec v^j+\tau^{ij}]+\partial_i(\delta\partial^i\phi)\,,
\ee
where the factor $1/(1+\delta)$ has cancelled, and $\phi$ is the gravitational potential. The two derivatives acting on $\tau^{ij}$ ensure the required scaling with wavenumber squared when going to Fourier space\footnote{Note that $\partial_i(\delta\partial^i\phi)=\partial_i\partial_j(\partial^i\phi\partial^j\phi-\phi(\partial^i\partial^j\phi-\delta^{ij}\partial^2\phi))/(3\Omega_\text{m}{\cal H}^2)$.}. We stress that this property holds independent of the assumed form of $\tau^{ij}$.

 The EFT expansion of the effective stress tensor $\tau^{ij}$ has the schematic form 
\be
  \tau^{ij} = \sum_{\cal O} {c_{\cal O}}\times {\cal O}\,,
\ee
where the sum runs over effective ``operators'' ${\cal O}$, being products of the coarse-grained fields, multiplied by prefactors $c_{\cal O}$, known as Wilson coefficients. The expansion is in close analogy to the generalized bias expansion, with the Wilson coefficients taking the role of the bias parameters. Within the large-scale structure literature, it is customary to denote the Wilson coefficients as ``counterterms'', despite the fact that they capture the (finite) impact of non-perturbative small-scale dynamics on the evolution at large scales, in addition to correcting for the spurious UV sensitivity of SPT. We refer to them as \emph{EFT parameters}.

The form of the operators is constrained by the symmetries of the system~\cite{Mercolli:2013bsa,Abolhasani:2015mra}. They are composed of the elementary Galilean-invariant \emph{building blocks}
\be
  \partial^i\partial^j\Phi,\quad \partial^i\vec u^j\,,
\ee
where $\Phi=2\phi/(3\Omega_\text{m}{\cal H}^2)$ is the rescaled gravitational potential satisfying $\Delta\Phi=\delta$, and $\vec u=\vec v/(-{\cal H}f)$, where $f=d\ln D_1/d\ln a$ and $D_1$ is the linear growth factor. Taking a trace over the tensor indices yields the density contrast and rescaled velocity divergence, respectively, while the traceless projection is related to the tidal tensor
\be
  s^{ij}=(\partial^i\partial^j-\delta^{ij}\partial^2/3)\Phi\,,
\ee
and velocity shear tensor 
\be
  \eta^{ij}=\partial^i\vec u^j+\partial^j\vec u^i-\frac23\delta^{ij}\nabla\cdot\vec u\,,
\ee  
respectively.

Since perturbations both below and above the coarse-graining scale evolve on comparable time-scales, the EFT is in general non-local in time~\cite{Abolhasani:2015mra}. In particular, the effective stress tensor can depend on the entire past evolution along the Lagrangian trajectory $\vec x_\text{fl}(\tau;\tau_0,\vec x_0)$ of a fluid element that is at position $\vec x_0$ at time $\tau_0$, i.e.\ satisfies the boundary condition $\vec x_\text{fl}(\tau_0;\tau_0,\vec x_0)=\vec x_0$ and $d\vec x_\text{fl}/d\tau=\vec v(\tau,\vec x_\text{fl}(\tau;\tau_0,\vec x_0))$. In addition, since the operators depend on products of the elementary building blocks, one needs to integrate over the past trajectory for each of the building blocks. The most general structure of the EFT expansion therefore takes the form
\be\label{eq:tauijnonlocal}
  \tau^{ij} = \sum_{\cal O} \int^\tau d\tau_1\cdots d\tau_{n_{\cal O}}\, {c_{\cal O}(\tau,\{\tau_k\})} {\cal O}(\tau,\vec x,\{\tau_k\}),\qquad 
  {\cal O}(\tau,\vec x,\{\tau_k\})\equiv \prod_{k=1}^{n_{\cal O}} {\cal O}_k(\tau_k,\vec x_\text{fl}(\tau_k;\tau,\vec x))\,,
\ee
where each ${\cal O}_k$ is given by spatial gradients of either $\partial^i\partial^j\Phi$ or $\partial^i\vec u^j$, respectively. Furthermore,  all possibilities to combine the vector indices among the ${\cal O}_k$ need to be  treated as possible operators that may appear in the sum over all ${\cal O}$.

The operators ${\cal O}$ can be classified by the lowest order in perturbation theory at which they contribute, as well as the total number of spatial gradients contained in them. In Fourier space, operators with a higher number of gradients are suppressed by higher powers of $k/\Lambda$, where $k$ is the typical magnitude of external wavenumbers that the observables of interest depend on, and $\Lambda$ is related to the non-linear scale.
The effective theory therefore provides a systematic method to compute corrections to SPT in a power series in $k/\Lambda$ by taking all possible operators up to a given order into account. 

Note that there are other well-known examples of effective theories that are non-local, for example soft collinear effective theory (SCET) \cite{Bauer:2000yr,Beneke:2002ph}, that is non-local along the light-cone direction of energetic particles involved in high-energy particle collisions, and possesses a somewhat analogous EFT expansion, see e.g.~\cite{Beneke:2017ztn}.

The operators with $n_{\cal O}=1$ and at lowest order in gradients are
\be
  {\cal O}\supset\{\delta^{ij}\delta,s^{ij},\delta^{ij}\nabla\cdot\vec u, \eta^{ij}\}\,,
\ee
with $\delta^{ij}$ being the Kronecker symbol. 
The associated Wilson coefficients correspond to the effective pressure, anisotropic stress, bulk- and shear viscosity, respectively, generalized to a non-local time-dependence. Operators with $n_{\cal O}=2$ appear starting at second order,
\be
  {\cal O}\supset\{\delta^{ij}\delta_{\tau_1}\delta_{\tau_2},s^{ij}_{\tau_1}\delta_{\tau_2},s^{ik}_{\tau_1}s^{kj}_{\tau_2},\delta^{ij} s^{lk}_{\tau_1}s^{kl}_{\tau_2}\}\,,
\ee
where the subscript denotes the argument, e.g.~$\delta_{\tau_k}=\delta(\tau_k,\vec x_\text{fl}(\tau_k;\tau,\vec x))$, and we omitted analogous terms involving velocity fields for brevity.

Using
\be 
  \frac{d}{d\tau_k}f(\tau_k,\vec x_\text{fl}(\tau_k;\tau,\vec x))\big|_{\tau_k=\tau}=(\partial_\tau+\vec v(\tau,\vec x)\cdot\nabla)f(\tau,\vec x)\,,
\ee
it is possible to Taylor expand each building block ${\cal O}_k$ around $\tau_k=\tau$. In particular,
\be
  {\cal O}_k(\tau_k,\vec x_\text{fl}(\tau_k;\tau,\vec x))
  = \sum_n \frac{1}{n!}(D_1(\tau_k)-D_1(\tau))^n\left[\frac{1}{{\cal H}fD_1(\tau)}(\partial_\tau+\vec v(\tau,\vec x)\cdot\nabla)\right]^n{\cal O}_k(\tau,\vec x)\,,
\ee
where $D_1(\tau)$ is the linear growth factor.
Inserting this Taylor expansion into \eqref{eq:tauijnonlocal} yields a local-in-time EFT expansion of $\tau^{ij}$. However, this comes at the price of introducing for each non-local-in-time operator an infinite tower of local-in-time operators with successively higher powers of convective time derivatives. Unlike spatial gradients, that are suppressed by powers of $k/\Lambda$, higher time derivatives yield factors of order one, such that higher-order terms in the Taylor expansion are in general not suppressed.
It is therefore necessary to keep the infinite series of higher-time-derivative operators, which is equivalent to adopting the non-local-in-time formulation.

Nevertheless, under certain conditions only a finite number of terms in the Taylor expansion is linearly independent when working at a finite order in perturbation theory. This happens in particular when assuming that the only source of time-dependence are powers of the linear growth factor $D_1$, as occurs for example in SPT in the EdS approximation. This property is also inherited by the EFT provided the EFT terms are introduced to correct for the UV sensitivity of SPT only. Assume for example that ${\cal O}_k$ can be written as a polynomial in $D_1(\tau)$ up to order $N$, when expanding to order $N$ in perturbation theory. Then only terms with $n\leq N$ need to be considered in the Taylor series\footnote{Note that $\frac{1}{{\cal H}fD_1(\tau)}(\partial_\tau+\vec v(\tau,\vec x)\cdot\nabla)=\frac{\partial}{\partial D_1(\tau)}-\frac{1}{D_1(\tau)}\frac{\vec v}{(-{\cal H}f)}\cdot\nabla$.}, leading to a finite set of local-in-time operators when working to a finite order in perturbation theory. Note that, since each occurrence of $\vec v$ adds at least one order, this argument applies also to the convective derivative when expanding the result up to terms of the desired order $N$. A convenient basis of operators can be constructed from the set of building blocks $\Pi^{[n]}$, defined recursively by~\cite{Mirbabayi:2014zca}
\be 
  \Pi^{[n]}=\frac{1}{(n-1)!}\left(\left[\frac{1}{{\cal H}f}(\partial_\tau+\vec v(\tau,\vec x)\cdot\nabla)\right]\Pi^{[n-1]}-(n-1)\Pi^{[n-1]}\right),\qquad \Pi^{[1]}= \partial^i\partial^j\Phi\,,
\ee
and an analogous set constructed starting from 
$\Pi^{[1]}_v = \partial^i\vec u^j$. The superscript indicates that the operator is of order $n$ in perturbation theory or higher. The operators ${\cal O}$ contributing to the local-in-time EFT expansion consist of products of these building blocks, and the expansion has the simplified form
\be\label{eq:tauijlocal}
  \tau^{ij} = \sum_{\cal O}  {c_{\cal O}(\tau)} {\cal O}(\tau,\vec x)\,.
\ee
At the first and second order, possible ${\cal O}$ are~\cite{Abolhasani:2015mra},
\bea 
\text{1st} &\quad& {\bf 1}\tr\Pi^{[1]}\,,\nn\\
\text{2nd} &\quad& {\bf 1}(\tr\Pi^{[1]})^2, \Pi^{[1]}\tr\Pi^{[1]}, {\bf 1}\tr(\Pi^{[1]})^2\,.
\eea
Here ${\bf 1}=\delta^{ij}$ and e.g.\ $\tr(\Pi^{[1]})^2=\Pi^{[1]kl}\Pi^{[1]lk}$.
We omitted analogous operators involving $\Pi_v^{[n]}$, which yield redundant contributions to $\tau_\theta$ at second order. In addition, the same applies to possible operators  $\Pi^{[1]}$, $(\Pi^{[1]})^2, \Pi^{[2]}$ and ${\bf 1}\tr(\Pi^{[2]})$~\cite{Mirbabayi:2014zca,Abolhasani:2015mra}.
In addition, the second order terms obtained from expanding $1/(1+\delta)$ in \eqref{eq:tautheta} are redundant at this order. Therefore there is one operator starting at first order, and three starting at second order. Using the freedom to choose a basis, we follow \cite{Baldauf:2014qfa} and use the following equivalent operator basis for $\tau_\theta$ at second order,
\be\label{eq:secondorderEFToperators}
  \Delta\delta,\quad \Delta\delta^2,\quad \Delta s^2,\quad \partial_i[s^{ij}\partial_j\delta]\,.
\ee
The four operators from above are sufficient to absorb the dominant UV sensitivity of the one-loop power and bispectrum, related to terms scaling as $k^2\int^\Lambda d^3q P_{11}(q)/q^2$, as we briefly review below. Here $k$ stands for an external wavenumber that is assumed to be in the BAO range, $q$ is the loop integration variable, and $P_{11}(q)$ is the usual linear power spectrum. In section\,\ref{sec:twoloop} we show that the four operators absorb also the dominant UV sensitivity of the two-loop bispectrum coming from the region where both loop wavenumbers are large, related to terms scaling as 
\be
  k^2\int^\Lambda d^3q_1\int^{q_1}d^3q_2 P_{11}(q_1)P_{11}(q_2)/q_1^2\,.
\ee
In addition, we discuss how to extend the EFT treatment to account for the region where only one of the two wavenumbers becomes large, related to ``nested divergences'' with an effective one-loop UV sensitivity $\propto \int^\Lambda d^3q P_{11}(q)/q^2$. In this work we do not take into account higher-gradient operators, that involve more spatial derivatives, and are suppressed by higher powers of $k/\Lambda$ relative to the dominant UV sensitive contributions given above. In addition, we omit contact terms involving (derivatives of) the Dirac delta $\delta_D^{(3)}(\vec x-\vec y)$ that arise when performing the EFT expansion for products of fields, e.g.~$\tau_\theta(\vec x)\tau_\theta(\vec y)\supset c_{\Delta}\Delta\delta_D^{(3)}(\vec x-\vec y)$, and are known as noise terms. They are irrelevant when focusing on the dominant UV sensitivity as defined above. Nevertheless, a classification of noise terms and an estimate of higher gradient contributions is presented in section\,\ref{sec:doublehard} and section\,\ref{sec:k4}, respectively.

\section{Review: One-loop bispectrum and EFT description}\label{sec:oneloop}

\subsection{One-loop SPT bispectrum}

We first discuss the SPT expressions for the one-loop bispectrum before discussing the corresponding EFT terms arising from the effective stress tensor.
The leading order contribution is the tree-level bispectrum induced by the quadratic coupling kernel $F^\text{(s)}_2$~\cite{Bernardeau:2001qr},
\be
B_\text{tree}(k_{1},k_{2},k_{3})= B_{211}(k_{1},k_{2},k_{3}) + 2\,\text{permutations}\,,
\ee
where
\be
B_{211}(k_{1},k_{2},k_{3})=2 F^\text{(s)}_2(\vk_2,\vk_3)P_{11}(k_2)P_{11}(k_3)\,,
\ee
and $P_{11}$ denotes the usual linear power spectrum. The one-loop contribution to the bispectrum is given by~\cite{Scoccimarro:1997st}
\be
  B_{1L} = B_{411}^s+B_{321}^s+B_{222}\,,
\ee
where
\begin{align}
&B_{411}(k_{1},k_{2},k_{3})      = 12 P_{11}(k_{2})P_{11}(k_{3})\int_{\vec q}
                                    F^\text{(s)}_{4}(\vk_{2}, \vk_{3}, \vq,-\vq)
                                    P_{11}(q)~,\\
&B_{321}^I(k_{1},k_{2},k_{3})    = 6 P_{11}(k_{3})\int_{\vec q}
                                    F^\text{(s)}_{3}(\vk_{3},\vk_{2}-\vq,\vq)
                                    F^\text{(s)}_{2}(-\vk_{2}+\vq,-\vq)
                                    P_{11}(q)P_{11}(|\vk_{2}-\vq|)~,\\
&B_{321}^{II}(k_{1},k_{2},k_{3}) = 6 P_{11}(k_{2})P_{11}(k_{3})\int_{\vec q}
                                    F^\text{(s)}_{3}(-\vk_{3}, \vq,-\vq)
                                    F^\text{(s)}_{2}(\vk_{2},\vk_{3})
                                    P_{11}(q)~,\\
&B_{222}(k_{1},k_{2},k_{3})      = 8 \int_{\vec q}
                                    F^\text{(s)}_{2}(-\vk_{3} - \vq, \vq)
                                    F^\text{(s)}_{2}(\vk_{3}+\vq, \vk_{2} - \vq))
                                    F^\text{(s)}_{2}(-\vk_{2}+\vq, - \vq)
                                    \nonumber\\
&\hspace{3cm}\times P_{11}(q)P_{11}(|\vk_{2}-\vq|)P_{11}(|\vk_{3}+\vq|)~,
\end{align}
and the symmetrized versions (denoted by the superscript $s$) are obtained by adding 2 permutations to $B_{411}$ and 5 permutations to $B_{321}\equiv B_{321}^I+B_{321}^{II}$, respectively. We use the shorthand notation $\int_{\vec{q}}\equiv \int d^3q$.

\subsection{Hard limit}\label{sec:hard1L}

The reason for introducing EFT parameters is that the contribution to the loop integrals from large wavenumbers are affected by strongly non-linear effects that cannot be treated perturbatively. Therefore, the EFT needs to be set up such as to correct for these uncertainties. In order to identify the corresponding terms that need to be added, it is instructive to investigate the asymptotic form of the SPT integrals for large loop wavenumber. We refer to the corresponding integration domain $q\gg k_1,k_2,k_3$ as \emph{hard} limit, for which the dependence of the SPT one-loop bispectrum on the external wavenumbers $k_i$ is given by~\cite{Baldauf:2014qfa}
\be\label{eq:B1Lhard}
  B_{1L}( k_1, k_2, k_3) \to B_{1L}^{h}( k_1, k_2, k_3)\sigmadsq(\Lambda) +{\cal O}(k^4P_L(k)^2,k^4P_L(k),k^6)\,,
\ee
where we expanded for $k_1/q\propto k_2/q\propto k_3/q\propto k/q\to 0$, and the displacement dispersion
\be
  \sigmadsq(\Lambda) = \frac13 \int_{q<\Lambda} d^3q \frac{P_{11}(q)}{q^2}\,,
\ee
captures the leading dependence of the loop integration on a UV cutoff $\Lambda$. In addition, 
\be
  B_{1L}^{h}( k_1, k_2, k_3) = b_{1L}^{h}(\vec k_2, \vec k_3)P_{11}(k_2)P_{11}(k_3)\ + 2\,\text{permutations}\,,
\ee
describes the dependence on the external wavenumbers, given by the one-loop SPT shape function
\be
  b_{1L}^{h}(\vec k,\vec p) = 3\left[12 f_4(\vec k,\vec p)-6\frac{61}{1890} f_3(\vec k,\vec p) \right]\,.
\ee
The two terms in the bracket correspond to the contributions from $B_{411}$ and $B_{321}^{II}$, respectively, based
on the asymptotic behaviour of the symmetrized SPT kernels for $q\to \infty$ \cite{Baldauf:2014qfa}
\bea
  \int \frac{d\Omega_q}{4\pi}\,F^\text{(s)}_3(\vec k,\vec q,-\vec q) &\to& -\frac{61}{1890}\frac{k^2}{q^2}+{\cal O}(k/q)^4\,,\\
  \int \frac{d\Omega_q}{4\pi}\,F^\text{(s)}_4(\vec k,\vec p,\vec q,-\vec q) &\to& \frac{1}{q^2}f_4(\vec k,\vec p)+{\cal O}({\rm max}(k,p)/q)^4\,,
\eea
where we defined the two shape functions
\bea
 f_3(\vec k,\vec p) &=& (k^2+p^2)F^\text{(s)}_2(\vec k,\vec p)\,,\nn\\
 f_4(\vec k,\vec p) &=& -\frac{49636 \mu^3 k p + 58812 (k^2 + p^2) + 
  114624 \mu^2 (k^2 + p^2) + 
  \mu k p \left(32879 \left(\frac{k^2}{p^2}+\frac{p^2}{k^2}\right) + 231478 \right)}{4074840 }\,,
\eea
with $\mu=\cos(\vec k,\vec p)$. Note that the hard limit of $B_{321}^{II}$ is related to the hard limit of the one-loop power spectrum, 
\begin{equation}\label{eq:P1Lhard}
    P_{1L}(k) = 2P_{13}+P_{22}\to P_{1L}^h(k) = - 2\times 3 \times 3\times \frac{61}{1890}\, k^2 \sigmadsq(\Lambda)  P_{11}(k)\,.
\end{equation}
The shape functions can be written as a linear combination 
\be
  f_i(\vec k,\vec p) = \sum_{j=1}^5 f_i^{(j)} \, b^{(j)}(\vec k,\vec p)\,,
\ee
of the basis shape functions
\bea\label{eq:shapes}
  b^{(1)}(\vec k,\vec p) &=& k^2+p^2\,,\nn\\
  b^{(2)}(\vec k,\vec p) &=& \vec k\cdot\vec p\,,\nn\\
  b^{(3)}(\vec k,\vec p) &=& \vec k\cdot\vec p \left(\frac{k^2}{p^2}+\frac{p^2}{k^2}\right)\,,\nn\\
  b^{(4)}(\vec k,\vec p) &=& \mu^2 (k^2+p^2)\,,\nn\\
  b^{(5)}(\vec k,\vec p) &=& \mu^3 k p\,.
\eea
For the numerical coefficients $f_i^{(j)}$ we refer to Tab.\,\ref{tab:shape} below.
Due to momentum conservation, it is well-known that the SPT kernels vanish when the sum of all arguments goes to zero. This implies that $f_i(\vec k,-\vec k)=0$, and leads to a condition on the coefficients,
\be\label{eq:shapeconstraint}
  2f_i^{(1)}-f_i^{(2)}-2f_i^{(3)}+2f_i^{(4)}-f_i^{(5)}=0\,,
\ee
which is indeed satisfied. Therefore, only four of the five coefficients are independent, and the hard limit can be completely specified by four parameters. The four independent shape functions precisely correspond to the EFT terms for the bispectrum given in \eqref{eq:secondorderEFToperators}, as we will review next.

\subsection{One-loop EFT terms}\label{sec:effstress}

The EFT modelling of the one-loop bispectrum was first presented in \cite{Baldauf:2014qfa,Angulo:2014tfa}. Here we follow the approach and notation presented in \cite{Baldauf:2014qfa} and \cite{Steele:2020tak}. The EFT contribution to the one-loop bispectrum can be written in the form
\bea
  B_{1L}^{\text{ctr}}(k_1,k_2,k_3) &=& B_{\tilde 211}(k_1,k_2,k_3) + 2 \ \text{permutations}\nn\\
  && {} + B_{ 2\tilde 11}(k_1,k_2,k_3) + 5 \ \text{permutations}
\eea
where
\bea
  B_{\tilde 211}(k_1,k_2,k_3) &=& 2\tilde F_2(\vk_2,\vk_3)P_{11}(k_2)P_{11}(k_3)\,,\nn\\
  B_{2\tilde 11}(k_1,k_2,k_3) &=& 2 F^\text{(s)}_2(\vk_2,\vk_3)P_{\tilde 11}(k_2)P_{11}(k_3)\,,\nn\\
\eea
with $P_{\tilde 1 1}=\tilde F_1 P_{11}(k)$ and
\be
  \tilde F_1=-\cssq k^2\,.
\ee
The EFT parameter $\cssq$ is known from the one-loop power spectrum,
\be\label{eq:P1Lctr}
  P_{1L}^{\text{ctr}} = 2P_{\tilde 11} = 2\tilde F_1 P_{11}=-2\cssq k^2 P_{11}\,,
\ee
and can be viewed to originate from a contribution to the effective stress tensor $\tau_\theta$ proportional to
\be
  \Delta \delta_1\,,
\ee
where $\delta_1$ is the linear density field. 

The kernel $\tilde F_2(\vk_2,\vk_3)$ can be written as a linear combination of the basis shape functions \eqref{eq:shapes}. As usual, the EFT enjoys the same symmetry as the underlying theory, and therefore momentum conservation requires that also $\tilde F_2(\vk,\vp)$ becomes zero for $\vk+\vp\to 0$. This means the linear combination of basis shape functions needs to satisfy the constraint \eqref{eq:shapeconstraint}, which implies that it can be fully specified by \emph{four} independent coefficients.
Following \cite{Steele:2020tak}, we parametrize these four coefficients by $\epsilon_{1,2,3}$ and $\gamma_{1-\text{loop}}$\footnote{The parameter $\gamma_{1-\text{loop}}$ is denoted by $\gamma_2$ in \cite{Steele:2020tak}, and the parameter $\cssq$ appearing in $B_{2\tilde 11}$ by $\gamma_1$.},
\begin{equation}
\tilde{F}_{2}(\vk_{1},\vk_{2})=-\left[\sum_{i=1}^{3}\epsilon_{i}E_{i}(\vk_{1},\vk_{2})+\gamma_{1-\text{loop}}\Gamma(\vk_{1},\vk_{2})\right]\, ,
\label{eq:symmf2tilde}
\end{equation}
where \cite{Steele:2020tak} 
\bea
E_1(\vec k_1,\vec k_2)&=&(\vec k_1+\vec k_2)^2~,\nn\\
E_2(\vec k_1,\vec k_2)&=&(\vec k_1+\vec k_2)^2\left(\frac{(\vec k_1\cdot \vec k_2)^2}{k_1^2k_2^2}-\frac13 \right)~,\nn\\
E_3(\vec k_1,\vec k_2)&=&-\frac{1}{6}(\vec k_{1}+\vec k_{2})^{2}+\frac{1}{2}\vk_{1}\cdot\vk_{2}\left(\frac{(\vk_{1}+\vk_{2})\cdot\vk_{2}}{k_{2}^{2}}+\frac{(\vk_{1}+\vk_{2})\cdot\vk_{1}}{k_{1}^{2}}\right)~,\nn\\
\Gamma(\vec k_1,\vec k_2)&=& (\vec k_1+\vec k_2)^2F^\text{(s)}_2(\vec k_1,\vec k_2)+\frac{2}{11}\left(\frac{10}{21}E_1(\vec k_1,\vec k_2)-\frac{5}{7}E_2(\vec k_1,\vec k_2)-3E_3(\vec k_1,\vec k_2)\right)\,.
\eea
The shape functions $E_1, E_2, E_3$  correspond to the kernels obtained from the second-order position space operators given in \eqref{eq:secondorderEFToperators},
\be
\Delta\delta_1^2,\quad \Delta s^2,\quad \partial_i[s^{ij}\partial_j\delta_1]\,,
\ee
respectively, where $s^{ij}$ is the tidal tensor, and $s^2=s^{ij}s_{ij}$. 
The kernel $\Gamma$ contains a combination of contributions from 
\be
  \Delta\delta_2\,,
\ee
where $\delta_2$ is the second-order density field, as well as second-order terms obtained when solving the continuity and Euler equations in presence of a term $\Delta \delta_1$ within the effective stress tensor\footnote{Here we evaluated $\Gamma$ for $m=1$ in the notation of \cite{Steele:2020tak}. A different choice of $m$ would lead to a different basis. Since it is spanning the same space of functions, one obtains identical results for the EFT bispectrum irrespective of $m$, when properly transforming the coefficients multiplying the basis functions as well. Also note that the operator $\Delta\theta_2$ is redundant, which can be seen using $(\vec k_1+\vec k_2)^2(G^\text{(s)}_2-F^\text{(s)}_2)=-\frac{4}{21}E_1+\frac27 E_2$.}.   Here and below we consider the EFT parameters as being free parameters at a given, fixed redshift of interest. One may view these parameters as being given by a convolution of the time-dependent terms multiplying the various contributions to the effective stress tensor with the linear propagator of the density field~\cite{Steele:2020tak}.

\medskip 

The most important property for our purposes is that the $E_1, E_2, E_3, \Gamma$ shapes form a basis of shape functions. Indeed, one realizes that it is related to the shape functions $b^{(1)}-b^{(5)}$ introduced in \eqref{eq:shapes} by a basis transformation, when eliminating one of the five shapes in \eqref{eq:shapes} by assuming the constraint \eqref{eq:shapeconstraint} from momentum conservation holds. Concretely, we have
\be
  \sum_{j=1}^5 f^{(j)} \, b^{(j)}(\vec k,\vec p) = \sum_{i=1}^{3} e_{i}E_{i}(\vk,\vp)+\gamma \Gamma(\vk,\vp)
\ee
with the mapping of coefficients given by
\bea
  f^{(1)} &=& e_1 - e_2/3 - e_3/6 + 72 \gamma/77\,,\nn\\
  f^{(2)} &=& 2e_1 - 2e_2/3 +2e_3/3 + 179 \gamma/77\,,\nn\\
  f^{(3)} &=&    \gamma/2\,,\nn\\
  f^{(4)} &=& e_2 + e_3/2 + 68 \gamma/77\,,\nn\\
  f^{(5)} &=& 2 e_2  + 24 \gamma/77\,.
\eea
Consequently, we note that the shape functions $f_{3,4}(\vk,\vp)$ defined above as well as the hard limit $b_{1L}^h(\vk,\vp)$ can equivalently be expanded in the $E_i/\Gamma$ basis. The corresponding coefficients are given in Tab.\,\ref{tab:shape}.

\subsubsection{Symmetry-based approach}

Within the symmetry-based approach \cite{Steele:2020tak}, we use four independent EFT parameters for the one-loop bispectrum, which can be taken to be $\epsilon_{1,2,3}$ and $\gamma_{1-\text{loop}}$, as well as one parameter $\cssq$ for the one-loop power spectrum. The EFT power- and bispectrum can be written as
\bea\label{eq:1Lren}
  P_{1L}^{\text{ren}}(k) &=& P_{11}+P_{1L}+P_{1L}^{\text{ctr}} = P_{11}+2P_{13}+P_{22}-2k^2c_s^2P_{11}\,,\nn\\
  B_{1L}^{\text{ren}}(k_1,k_2,k_3) &=& B_\text{tree}+B_{1L}+B_{1L}^{\text{ctr}}\,,
\eea
with $B_{1L}$ being the SPT one-loop contribution to the bispectrum, and $B_{1L}^{\text{ctr}}$ obtained from combining $B_{\tilde 211}$ and $B_{2\tilde 1 1}$. After noting that the latter can be expanded in the basis of shape functions as well, the EFT contribution can be brought into the form
\be\label{eq:1Lctr}
  B_{1L}^{\text{ctr}}(k_1,k_2,k_3) = - 2\left[\sum_{i=1}^{3}\hat \epsilon_{i}E_{i}(\vk_{1},\vk_{2})+\hat \gamma_{1-\text{loop}}\Gamma(\vk_{1},\vk_{2})\right]P_{11}(k_2)P_{11}(k_3) +2\ \text{permutations}
\ee
where 
\bea
  \hat \epsilon_1 &=& \epsilon_1 - \frac{97}{231}\cssq\,,\nn\\
  \hat \epsilon_2 &=& \epsilon_2 - \frac{12}{77}\cssq\,,\nn\\
  \hat \epsilon_3 &=& \epsilon_3 - \frac{68}{77}\cssq\,,\nn\\
  \hat \gamma_{1-\text{loop}} &=& \gamma_{1-\text{loop}} + \cssq\,.
\eea
We note that, since, for the symmetry-based approach, all four parameters are allowed to vary freely, one could equivalently use the shifted values (with a hat) as input parameters. This argument also makes it apparent that the one-loop EFT \emph{bispectrum} within the symmetry-based approach depends on exactly \emph{four} independent input parameters.
In addition, we note that the bispectrum $B_{1L}^{\text{ren}}$ is \emph{independent of the chosen value of $\cssq$} after fitting $\epsilon_i$ and $\gamma_{1-\text{loop}}$ to simulation or observation data.
In practice, throughout this work, we use the numerical value of $\cssq=\cssq|_{P_{1L}}$ determined by fitting  the one-loop power spectrum spectrum to simulation data, and treat the following parameters as free:
\be
  \{ \gamma_{1-\text{loop}},\epsilon_1,\epsilon_2,\epsilon_3 \} \qquad\qquad \text{1L\ symmetry-based approach.}
\ee
We will see below that the symmetry-based approach can be extended to two-loop order.

\subsubsection{UV-inspired approach}\label{sec:UVinsp}

The so-called UV-inspired approach makes additional assumptions on the relative contributions of the basis of shape functions, motivated by the linear combination in which they appear in the hard limit of the SPT one-loop bispectrum. In particular, the hard limit of $B_{411}$ is proportional to the shape function $f_4(\vk,\vp)$, which is a particular linear combination of the $E_i$ and $\Gamma$ basis functions. The UV-inspired approach assumes that it is sufficient to parametrize $B_{\tilde 2 11}$ by a single free EFT parameter, which can be taken to be $\gamma_{1-\text{loop}}$, such that $\tilde F_2(\vk,\vp)\propto \gamma_{1-\text{loop}} f_4(\vk,\vp)$. Noting that $f_4(\vk,\vp)$ can be written as a particular linear combination of the shape basis functions, this corresponds to fixing the $\epsilon_i$ parameters in terms of $\gamma_{1-\text{loop}}$ as 
\bea\label{eq:epsUVinspired}
  \epsilon_1 = \frac{3466}{14091}\gamma_{1-\text{loop}}\,,\quad 
  \epsilon_2 = \frac{7285}{32879}\gamma_{1-\text{loop}}\,,\quad 
  \epsilon_3 = \frac{41982}{32879}\gamma_{1-\text{loop}}\,.\quad 
\eea
Assuming that the value of $\cssq$ is fixed by the one-loop power spectrum, the UV-inspired approach therefore introduces only one additional free parameter, $\gamma_{1-\text{loop}}$, for the bispectrum. 

An even more restrictive ansatz can be obtained by noticing that, at one-loop order, all terms with leading UV sensitivity are in the hard limit proportional to the same moment of the linear power spectrum, given by $\sigmadsq$. One may then investigate the hypothesis that UV effects correct this moment in a universal manner in all one-loop contributions. If this is the case, the correction from small-scale non-linearities can be taken into account by correcting the contribution from the hard region according to the universal replacement
\be
  \sigmadsq \mapsto \sigmadsq + \Delta\sigmadsq
\ee
with a single unknown parameter $\Delta\sigmadsq$. The EFT correction term is in this approach obtained by formally replacing $\sigmadsq$ by $\Delta\sigmadsq$ in the hard limit of the  bispectrum $B_{1L}^h$ given in \eqref{eq:B1Lhard}, as well as in the hard limit of the power spectrum $P_{1L}^h$, see \eqref{eq:P1Lhard}. By comparing with the usual parametrization of the one-loop EFT correction $P_{1L}^\text{ctr}$ from \eqref{eq:P1Lctr}, one can identify
\be
  \Delta\sigmadsq \equiv \frac{61}{210}\cssq\,. 
\ee
Furthermore, from \eqref{eq:B1Lhard} it follows that in this zero-parameter approach the bispectrum EFT parameter is chosen to be 
\be
  \gamma_{1-\text{loop}}=\cssq\,.
\ee
Various combinations of possibilities for  determining the EFT terms in either the one- or zero-parameter UV-inspired approach or the symmetry-based approach have been discussed in detail for the one-loop bispectrum in \cite{Steele:2020tak}. Here we consider the following possibilities 
\bea
 \{ \gamma_{1-\text{loop}} \} &\qquad\qquad& \text{1L\ UV-inspired 1-parameter with }\ \epsilon_i/\gamma_{1-\text{loop}}\ \text{fixed}\,, \\
 \emptyset &\qquad\qquad& \text{1L\ UV-inspired 0-parameter with}\ \epsilon_i/\gamma_{1-\text{loop}}\ \text{fixed and}\ \gamma_{1-\text{loop}}=\cssq|_{P_{1L}}\,.
\eea
As we will see, and as may be expected, the UV-inspired approach can strictly speaking not be extended to the two-loop bispectrum. Nevertheless, we will consider a naive extension and compare it to the symmetry-based approach at two-loop order further below.

\section{$\Lambda$CDM versus EdS time-dependence}\label{sec:exactoneloop}

Before discussing the two-loop bispectrum, we scrutinize the commonly adopted EdS approximation for the non-linear kernels. The exact time-dependence within $\Lambda$CDM can be taken into account by replacing the usual SPT-EdS kernels $F_n$ and $G_n$ by time-dependent kernels that obey the set of coupled, ordinary differential equations
\bea\label{kernels}
  (\partial_{\eta}+n)F_{n}(q_1,\dots ,q_n,\eta)-G_{n}(q_1,\dots ,q_n,\eta)
  &=& \sum_{m=1}^n \alpha(q_1+\cdots+q_m,q_{m+1}+\cdots+q_n) \nn\\
  && \times G_{m}(q_1,\dots,q_m,\eta)F_{n-m}(q_{m+1},\dots,q_n,\eta)\,, \nn\\
    (\partial_{\eta}+n+x(\eta)-1)G_{n}(q_1,\dots ,q_n,\eta)-x(\eta)F_{n}(q_1,\dots ,q_n,\eta)
  &=& \sum_{m=1}^n \beta(q_1+\cdots+q_m,q_{m+1}+\cdots+q_n) \nn\\
  && \times G_{m}(q_1,\dots,q_m,\eta)G_{n-m}(q_{m+1},\dots,q_n,\eta)\,,
\eea
where $x(\eta)\equiv \frac{3}{2}\frac{\Omega_m}{f^2}$, $\eta=\ln(D_1)$, $f=d\ln(D_1)/d\ln a$,  $\Omega_m=\Omega_{m0}a^{-3}H_0^2/H^2$, and $D_1$ is the usual linear growth factor. The kernels still need to be symmetrized with respect to arbitrary permutations of the wavenumbers. The conventional EdS approximation is recovered by approximating $x\mapsto \frac32$, in which case the equations possess an analytic solution with time-independent kernels. Indeed, the usual algebraic EdS-SPT recursion relation is obtained by setting $\partial_{\eta}\mapsto 0$,  $x\mapsto \frac32$ and taking suitable linear combinations of both equations.

Even though $x(\eta)$ deviates from $\frac32$ by about $15\%$ at $z=0$ for a realistic $\Lambda$CDM cosmology, the EdS approximation is known to work relatively well for the one-loop matter power spectrum \cite{Takahashi:2008yk,Fasiello:2016qpn, Garny:2020ilv, Donath:2020abv}. One reason is that $x$-dependent terms enter only via the decaying mode, whose contribution to the kernels is non-negligible but somewhat suppressed.

Analytic solutions to~\eqref{kernels} for general $x(\eta)$ have been derived in the literature up to third order~\cite{Bernardeau:2001qr,Takahashi:2008yk,Fasiello:2016qpn, Donath:2020abv}, in which the time- and wavenumber-dependence of the kernels are factorized. For e.g.\ the $F_3$ kernel, the time-dependence can be captured by three parameters $\nu_2$, $\nu_3$ and $\lambda_3$ each multiplying a wavenumber-dependent function~\cite{Bernardeau:2001qr}. The factorization implies that given a cosmology with a non-trivial $x(\eta)$, one only needs to determine the parameters in order to obtain the kernels. Nonetheless, solving~\eqref{kernels} analytically becomes increasingly more difficult at higher orders due to the recursive structure of the equations.

The approach of~\cite{Garny:2020ilv} is instead to solve the set of equations~\eqref{kernels} numerically for each configuration of wavevectors that is needed in the analysis. While this method entails numerous integrations of the kernel equations, it is easily extended to higher orders. In particular, the two-loop power spectrum (involving $F_5$) with exact time-dependence was computed in~\cite{Garny:2020ilv}, finding a deviation of $1\%$ at $k = 0.2\ihMpc$ compared to the EdS approximation. 

\begin{figure}
\includegraphics[width=0.99\textwidth]{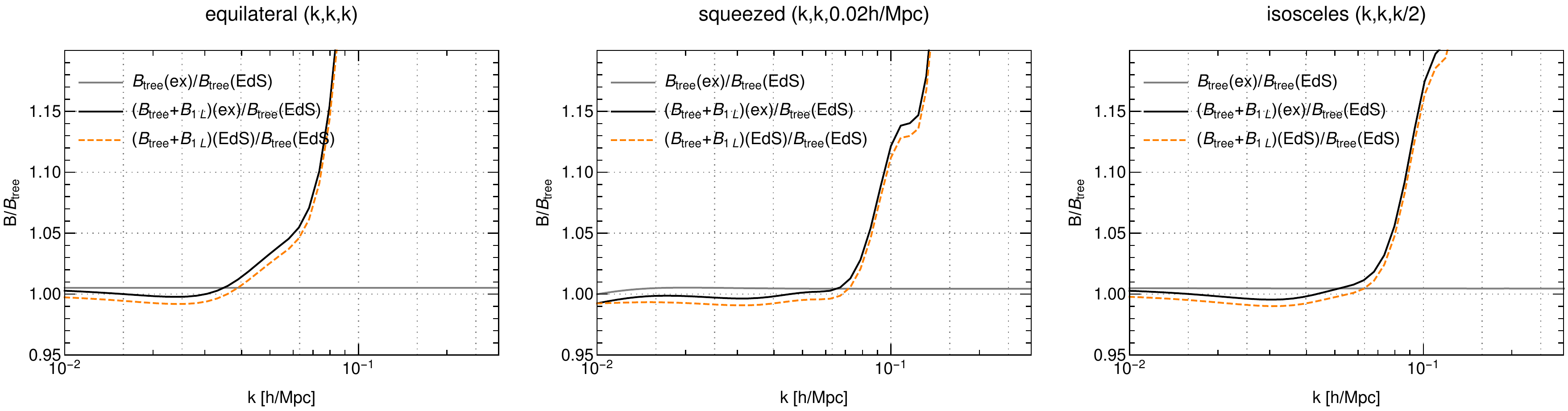}
\caption{\label{fig:Bex}
Comparison of the one-loop bispectrum computed with EdS kernels (orange dotted lines) with the case taking the exact time-dependence into account (black lines). Note that the exact kernels are taken into account both in the tree-level as well as the one-loop piece. For comparison we show the impact at tree-level only in gray. All lines are normalized to the tree-level EdS bispectrum, and we show three different shapes as indicated above the panels.
}
\end{figure}

In order to assess the impact of using EdS kernels we compute the one-loop bispectrum with exact time-dependence following the method developed in~\cite{Garny:2020ilv}. The result is shown in Fig.\,\ref{fig:Bex} for three different shapes (black lines), as compared to the EdS approximation (orange dotted lines). Apart from an approximately $k$-independent shift, that is due to the impact of exact time-dependence at tree-level, there is a $k$-dependent difference arising from the one-loop contribution. The relative size of tree and loop-contributions is shown in Fig.\,\ref{fig:Bex2}. The $k$-independent offset at tree-level is of order $0.5\%$, while the difference between exact and EdS results at one-loop increases with $k$, and crosses the $1\%$ threshold at $k\simeq 0.1\ihMpc$ for the shapes considered here.

\subsection{Degeneracy with the EFT parameters}

\begin{figure}
\includegraphics[width=0.99\textwidth]{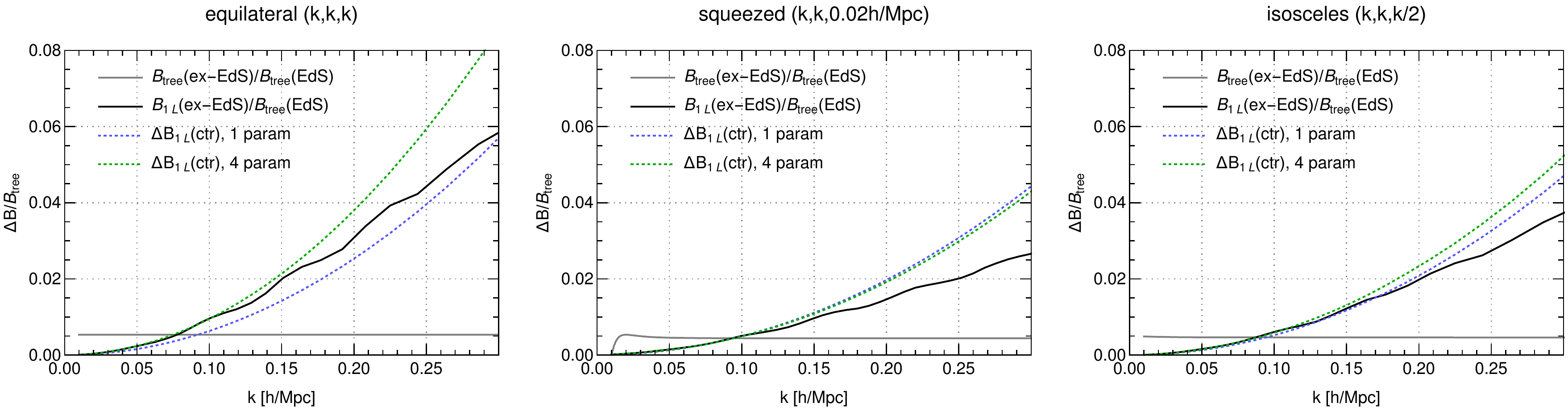}
\caption{\label{fig:Bex2}
Relative change of the one-loop contribution to the bispectrum computed with exact or EdS kernels  (black lines), as well as the best-fit EFT contribution \eqref{eq:1Lctr} to the bispectrum in the UV-inspired approach (blue lines) or symmetry-based approach (green lines). Gray points show the relative difference between exact and EdS kernels at tree-level for comparison.
}
\end{figure}

Within the effective theory approach, one may ask to which extent the difference between the bispectrum obtained in the EdS approximation and for the exact time-dependence can be absorbed by a shift in the EFT parameters. Clearly, the impact of exact time-dependence on the \emph{tree-level} bispectrum is not degenerate with EFT parameters, since the former yields a difference that persists even on large scales, while the EFT contributions are suppressed by factors of order $k^2$ in that limit. However, since the exact $F_2$ kernel can easily be implemented for the tree-level spectrum, the question that is in practice more relevant is the extent to which the error made by the EdS approximation is degenerate with EFT parameters at the loop level.

To illustrate this point, we first review the case of the one-loop power spectrum. For $k\to 0$, it scales as
\begin{equation}
    P_{1L}(k) = 2P_{13}+P_{22}\to - 2 c\, k^2 \sigma_d^2  P_{11}\,,
\end{equation}
and furthermore $P_{13}$ dominates in that limit. These properties are not changed when using the exact kernels, as $F_3(k,q,-q)\propto k^2/q^2$ for $k\to 0$ is guaranteed by momentum conservation arguments. However, the value of the proportionality constant $c \equiv \lim_{k\to 0}9\int \frac{d\Omega_q}{4\pi}\,F_3(k,q,-q)\frac{q^2}{k^2}$ changes. In the EdS approximation it is given by $c_{\text{EdS}}=61/210\simeq 0.2905$. When using the parametrization of the exact $F_3$ kernel from \cite{Bernardeau:2001qr} one finds
\be
 c = \frac{1}{30} \left(-121 - 16 \lambda_3 + 162 \nu_2 - 36 \nu_3\right)\,,
\ee
where $\nu_2,\nu_3$ and $\lambda_3$ are related to generalized second and third order growth functions \cite{Bernardeau:2001qr}. Their values in the EdS approximation are $34/21\simeq 1.6191, 682/189\simeq 3.6085, 1/6\simeq 0.1667$. For a  $\Lambda$CDM model (with $\Omega_{m0}=0.272$) one finds $1.6217, 3.6233, 0.1700$,     respectively at $z=0$, giving $c_{\text{$\Lambda$CDM}}=0.2853$. 
Therefore, for $k\to 0$,
\be
    P_\text{1L,$\Lambda$CDM}-P_\text{1L,EdS}\to -2(c_\text{$\Lambda$CDM}-c_{\text{EdS}})\sigmadsq  k^2 P_{11}\,.
    \ee
Thus, in the large-scale limit, the difference between approximate EdS and correct LCDM growth factors can be absorbed into a change of the speed-of-sound of the leading power spectrum EFT parameter,
\be
      \Delta c_s^2 = (c_\text{$\Lambda$CDM}-c_{\text{EdS}})\sigmadsq\,.
\ee
For the cosmology under consideration, this corresponds to a shift of about $\Delta \cssq\approx -0.2 \hMpcsq$, which is about $20\%$ of the total amplitude of the speed-of-sound term (see also \cite{Fasiello:2016qpn}).
    
For the bispectrum, on large scales the leading order difference between the exact $\Lambda$CDM and approximate EdS time-dependence is given by $B_{211}$. This difference is at the level of $k^0$ and and thus cannot be absorbed by the one-loop EFT parameters.\footnote{However, the tree-level correction is somewhat degenerate with the numerical correction factor $(\Delta D_2+2\Delta D_1) B_{211}$, see section~\ref{sec:numerics}.}
    
At the loop level, the growth factor corrections can be absorbed into a change of the EFT parameters in the limit of large scales. In Fig.\,\ref{fig:Bex2} we show the relative change of the bispectrum when using exact $\Lambda$CDM time-dependence versus the EdS approximation. In addition, we show by the blue and green dotted lines the part that can be absorbed into a shift of the EFT parameters, for the UV-inspired as well as symmetry-based approaches, respectively.  For the UV-inspired ansatz the shift in the $\gamma_{1-\text{loop}}$ parameter is similar to the one observed for $\cssq$ in the power spectrum. We find  that most, but not all of the difference can be absorbed by the EFT terms. In addition, it is apparent that the modification of $B_{211}$ (gray points) cannot be absorbed by EFT corrections, in line with the theoretical expectation.
    
\section{The two-loop bispectrum}\label{sec:twoloop}

\begin{figure}
    \centering
    \includegraphics[width=0.9\textwidth]{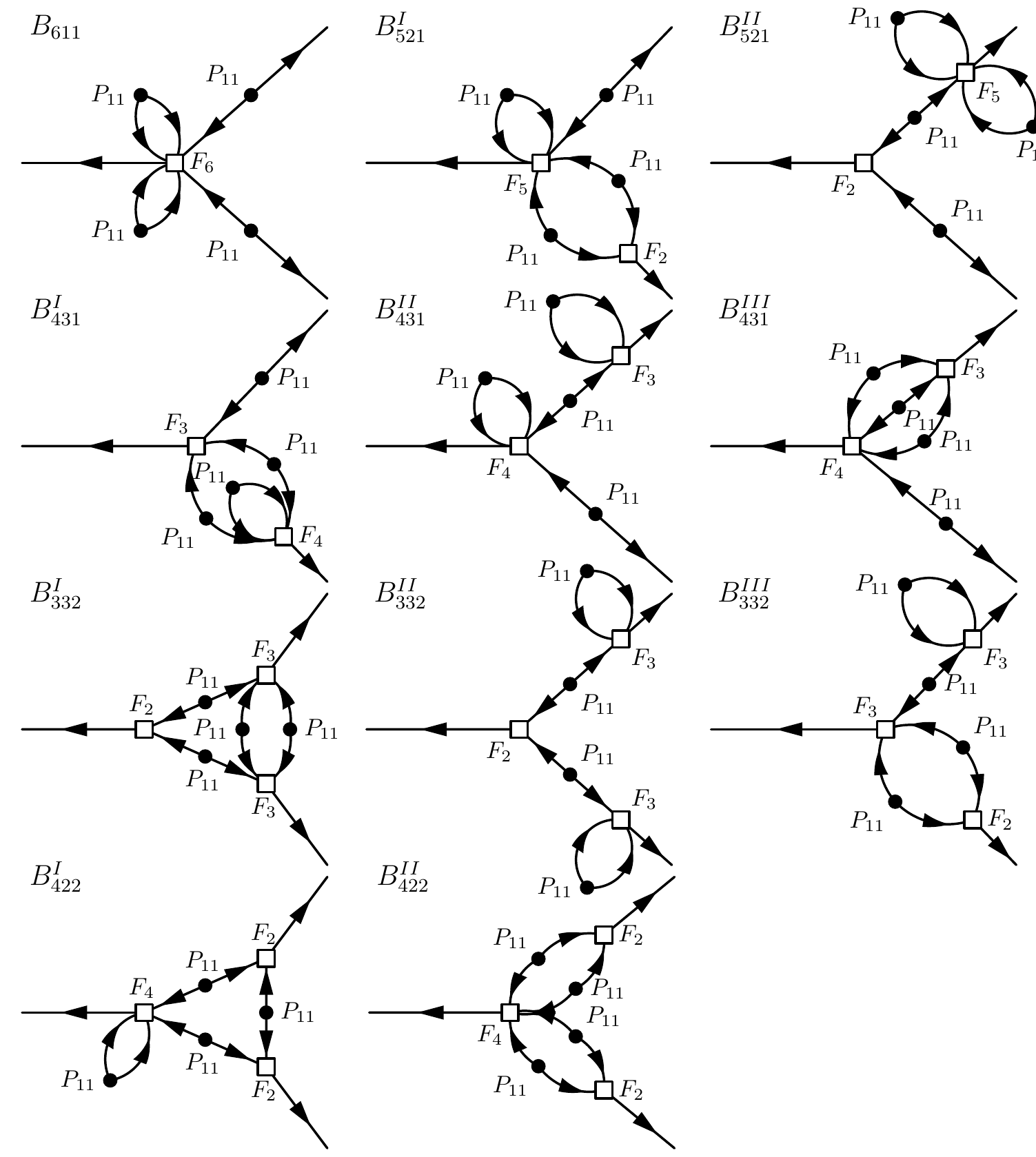}
    \caption{Diagrams of the two-loop bispectrum contributions. The daisy loops, i.e.\ loops that are only connected to a single $F_n$ kernel, are the leading sources of UV sensitivity of the loop.}
    \label{fig:diagrams}
\end{figure}

\subsection{Two-loop SPT bispectrum}

\begin{figure}
\includegraphics[width=0.99\textwidth]{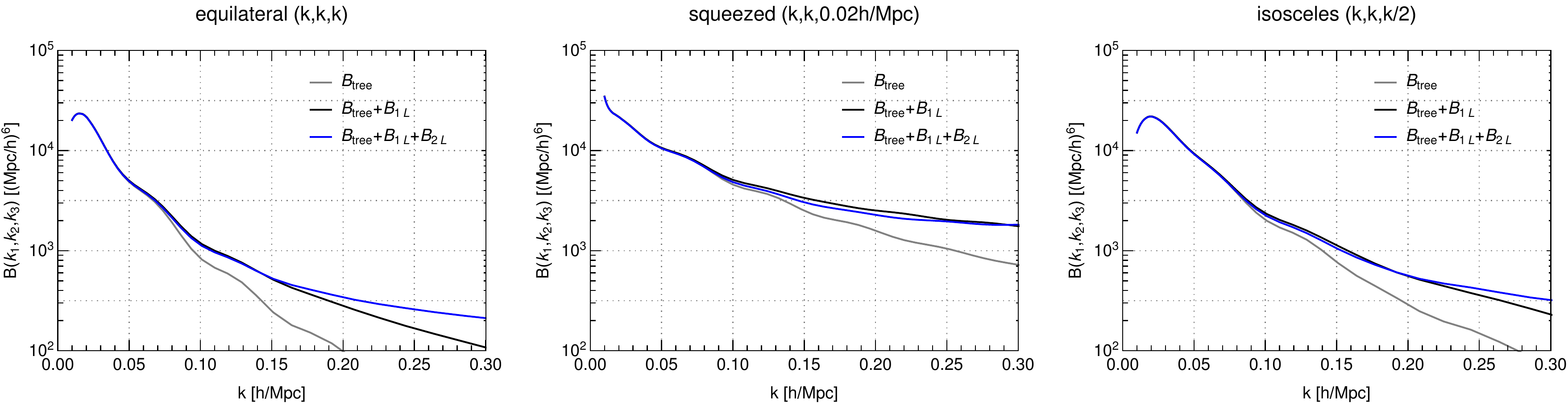}
\caption{\label{fig:B2L_spt_shapes}
Unrenormalized tree-level, one- and two-loop approximation to the bispectrum in SPT with exact time-dependence for the tree and one-loop contribution, and EdS-SPT kernels for the two-loop. All loop integrals are cut off at $0.6\ihMpc$ (the cutoff dependence is removed by renormalization, see below).
}
\end{figure}

Up to two loops, the equal time bispectrum within SPT is given by
\be
B_{\text{SPT}}=B_\text{tree}+B_{1L}+B_{2L}\,,
\ee
where
\be
  B_{2L}=B_{611}^s+B_{521}^s+B_{422}^s+B_{431}^s+B_{332}^s\,.
\ee
The diagrams for the two-loop SPT bispectrum are shown in Fig.~\ref{fig:diagrams} and correspond to the expressions 
\begin{align}\label{eq:B2Lexpressions}
    &B_{611}(k_{1},k_{2},k_{3})=90P_{11}(k_{2})P_{11}(k_{3}) \int_{\vq_1,\vq_2}F^\text{(s)}_{6}(\vk_{2},\vk_{3},\vq_{1},-\vq_{1},\vq_{2},-\vq_{2})P_{11}(q_{1})P_{11}(q_{2})\,,\nn\\
    &B_{521}^{I}(k_{1},k_{2},k_{3})=60P_{11}(k_{2})\int_{\vq_1,\vq_2}F^\text{(s)}_{5}(\vk_{2},\vq_{1},-\vq_{1},\vq_{2},-\vq_{2}+\vk_{3})F^\text{(s)}_{2}(-\vq_{2},\vq_{2}-\vk_{3})P_{11}(q_{1})P_{11}(q_{2})P_{11}(|\vq_{2}-\vk_{3}|)\,,\nn\\
    &B_{521}^{II}(k_{1},k_{2},k_{3})=30P_{11}(k_{2})P_{11}(k_{3})\int_{\vq_1,\vq_2}F^\text{(s)}_{5}(-\vk_{2},\vq_{1},-\vq_{1},\vq_{2},-\vq_{2})F^\text{(s)}_{2}(\vk_{2},\vk_{3})P_{11}(q_{1})P_{11}(q_{2})\,,\nn\\
    &B_{431}^{I}(k_{1},k_{2},k_{3})=36P_{11}(k_{3})\int_{\vq_1,\vq_2}F^\text{(s)}_{4}(-\vq_{1},\vq_{1}-\vk_{1},\vq_{2},-\vq_{2})F^\text{(s)}_{3}(\vk_{3},\vq_{1},-\vq_{1}+\vk_{1})P_{11}(q_{1})P_{11}(q_{2})P_{11}(|\vq_{1}-\vk_{1}|)\,,\nn\\
    &B_{431}^{II}(k_{1},k_{2},k_{3})=36P_{11}(k_{2})P_{11}(k_{3})\int_{\vq_1,\vq_2}F^\text{(s)}_{4}(\vk_{2},\vk_{3},\vq_{1},-\vq_{1})F^\text{(s)}_{3}(-\vk_{2},\vq_{2},-\vq_{2})P_{11}(q_{1})P_{11}(q_{2})\,,\nn\\
    &B_{431}^{III}(k_{1},k_{2},k_{3})=24P_{11}(k_{3})\int_{\vq_1,\vq_2}F^\text{(s)}_{4}(\vk_{2}+\vq_{1}+\vq_{2},\vk_{3},-\vq_{1},-\vq_{2})F^\text{(s)}_{3}(-\vk_{2}-\vq_{1}-\vq_{2},\vq_{1},\vq_{2})
    \nn\\&\hspace{3cm}\times P_{11}(q_{1})P_{11}(q_{2})P_{11}(|\vk_{2}-\vq_{1}-\vq_{2})\,,\nn\\
    &B_{332}^I(k_{1},k_{2},k_{3})= 36 \int_{\vq_1,\vq_2}
    F^\text{(s)}_{3}(\vq_{1},\vq_{2},-\vk_{1}-\vq_{1}-\vq_2)
    F^\text{(s)}_{3}(-\vq_{2},\vk_{1} + \vq_{1}+ \vq_2,\vk_{3}-\vq_1)
    F^\text{(s)}_{2}(-\vk_{3} + \vq_1,-\vq_{1})\nn\\&\hspace{3cm}\times P_{11}(q_{1})P_{11}(q_{2})P_{11}(|\vk_{1}+\vq_{1} + \vq_2|) P_{11}(|\vk_3 - \vq_1|)\nn\\
    &B_{332}^{II}(k_{1},k_{2},k_{3})=18P_{11}(k_{1})P_{11}(k_{2})\int_{\vq_1,\vq_2}F^\text{(s)}_{3}(-\vk_{1},\vq_{1},-\vq_{1})F^\text{(s)}_{3}(-\vk_{2},\vq_{2},-\vq_{2})F^\text{(s)}_{2}(\vk_{1},\vk_{2})P_{11}(q_{1})P_{11}(q_{2})\,,\nn\\
    &B_{332}^{III}(k_{1},k_{2},k_{3})=18P_{11}(k_{1})\int_{\vq_1,\vq_2}F^\text{(s)}_{3}(-\vk_{1},\vq_{1},-\vq_{1})F^\text{(s)}_{3}(\vk_{1},\vk_{3}+\vq_{2},-\vq_{2})F^\text{(s)}_{2}(\vq_{2},-\vq_{2}-\vk_{3})\nonumber\\&\hspace{3cm}\times P_{11}(q_{1})P_{11}(q_{2})P_{11}(|\vq_{2}+\vk_{3}|)\,,\nn\\
    &B_{422}^{I}(k_{1},k_{2},k_{3})=48\int_{\vq_1,\vq_2}F^\text{(s)}_{4}(\vq_{1},-\vq_{1},\vq_{2},-\vq_{2}+\vk_{2}+\vk_{3})F^\text{(s)}_{2}(-\vq_{2},\vq_{2}-\vk_{2})F^\text{(s)}_{2}(-\vq_{2}+\vk_{2},\vq_{2}-\vk_{2}-\vk_{3})\nonumber\\&\hspace{3cm}\times P_{11}(q_{1})P_{11}(q_{2})P_{11}(|\vq_{2}-\vk_{2}|)P_{11}(|\vq_{2}-\vk_{2}-\vk_{3}|)\,,\nn\\
    &B_{422}^{II}(k_{1},k_{2},k_{3})=48\int_{\vq_1,\vq_2}F^\text{(s)}_{4}(\vq_{1},\vq_{2},\vk_{2}-\vq_{1},\vk_{3}-\vq_{2})F^\text{(s)}_{2}(-\vq_{1},\vq_{1}-\vk_{2})F^\text{(s)}_{2}(-\vq_{2},\vq_{2}-\vk_{3})\nonumber\\&\hspace{3cm}\times P_{11}(q_{1})P_{11}(q_{2})P_{11}(|\vq_{1}-\vk_{2}|)P_{11}(|\vq_{2}-\vk_{3}|)\,.
\end{align}
The symmetrized expressions  are obtained by 
\bea
  B_{611}^s &=& B_{611} +  2\ \text{permutations}\,,\nn\\
  B_{521}^s &=& (B_{521}^I+B_{521}^{II}) +  5\ \text{permutations}\,,\nn\\
  B_{431}^s &=& (B_{431}^I+B_{431}^{II}+B_{431}^{III}) +  5\ \text{permutations}\,,\nn\\
  B_{332}^s &=& \left[ (B_{332}^I+B_{332}^{II}) +  2\ \text{permutations}\right] + \left[  B_{332}^{III}  +  5\ \text{permutations}\right]\,,\nn\\
  B_{422}^s &=& (B_{422}^I+B_{422}^{II}) +  2\ \text{permutations}\,.
\eea
For the computation of the two-loop bispectrum we follow the algorithm outlined in the appendix of~\cite{Floerchinger:2019eoj} (see also \cite{Lazanu:2018yae}). This in particular encompasses a suitable combination of all individual contributions, which makes sure that all leading and subleading infrared-enhanced contributions cancel  at the \emph{integrand} level. The numerical integration is then performed using the Monte Carlo integration package CUBA~\cite{Hahn:2004fe}. We performed numerous conversion tests and checked the result based on two independent implementations. Furthermore, we also compute the two-loop based on the grid-PT technique, for a given realization of the density field (see below).

The SPT result for the (unrenormalized) two-loop bispectrum is shown in Fig.\,\ref{fig:B2L_spt_shapes} for three different shapes. As expected, the result depends on contributions from large wavenumbers. The EFT program is precisely designed to take care of this UV dependence and encapsulate the uncertainties in EFT parameters. Before describing this procedure for the two-loop bispectrum, we briefly review the case of the power spectrum.

\subsection{Brief review of the EFT for the two-loop power spectrum}

At two-loop order, one has to discriminate two types of UV limits: the \emph{single-hard} limit with one of the two loop wavenumbers being large compared to the external wavenumber, and the \emph{double-hard} limit, where both loop wavenumbers become large compared to the external scales.
In \cite{Baldauf:2015aha} it was pointed out that in the two-loop power spectrum the double-hard contributions lead to a $k^2 P_{11}(k)$ contribution that is degenerate with the one-loop EFT parameter $\cssq k^2 P_{11}(k)$, whereas the single-hard contributions require new EFT parameters beyond the one-loop power spectrum. The leading single-hard contributions are proportional to the displacement dispersion $\sigmadsq$ in the $k\ll q$ limit, and arise from daisy diagrams. This analogy to the hard limit of the $P_{13}$ contribution was explored in \cite{Baldauf:2015aha} to propose a one-parameter ansatz for the EFT parameter, in which the degenerate double-hard limits of the two-loop expressions were absorbed into the leading $\cssq k^2 P_{11}(k)$ EFT correction, and the single-hard limit proportional to $\sigmadsq$ is used as additional EFT parameter upon replacing $\sigmadsq\mapsto \frac{210}{61}\cssq$ (see section~\ref{sec:oneloop}).

For the two-loop bispectrum a straightforward application of this procedure fails because the double-hard limit of the two-loop bispectrum is \emph{not} proportional to the hard limit of the one-loop bispectrum, i.e.\ the hard regions lead to a different shape dependence at one- and two-loop respectively. We will present the analysis of the double-hard region next, and then discuss the single-hard limit as well as a generalization of the EFT approach to the two-loop bispectrum.

\subsection{Double-hard limit}\label{sec:doublehard}

The EFT description at two-loop order requires to consider the limit of the SPT two-loop expressions when either both or one of the loop wavenumbers tend to infinity, denoted as double and single hard limit, respectively.
Below we present \emph{analytical} results for the double-hard limit, and discuss their properties. 

We consider the limit where both loop wavenumbers tend to infinity such that their ratio remains finite, $q_1\propto q_2\propto q\to \infty$, for fixed external wavenumbers $k_i$. Equivalently, this correspond to the limit $k_1\propto k_2\propto k_3\propto k\to 0$ with fixed $q_i$. In order to discuss the parametric dependence in this limit, we denote the scale of the loop wavenumbers by $q$, and the scale of the external wavenumbers by $k$. Due to momentum conservation, the symmetrized SPT kernels satisfy the decoupling property~\cite{Goroff:1986ep}
\be\label{eq:Fnlimit}
  F_n^{\text{(s)}} \sim k^2/q^2,\qquad k/q\to 0\,,
\ee
if the \emph{sum} of all arguments scales as $k\propto k_i$, while (a subset of) the arguments of the kernels scales parametrically as $q\propto q_i$, such as for example for $F^\text{(s)}_6(\vec k_1,\vec k_2,\vec q_1,-\vec q_1,\vec q_2,-\vec q_2)$. The relation above can be used to estimate the parametric scaling, while the precise expression involves an in general complicated dependence on the ratios $k_i/k_j$ (or equivalently on $k_1$, $k_2$ and the angle between the corresponding wavevectors) as well as on $q_1/q_2$, see below. For now, we use \eqref{eq:Fnlimit} to estimate the parametric scaling in the double-hard limit, which is schematically given by
\bea
  B_{611}^{hh} &\sim& k^2P_{11}(k_2)P_{11}(k_3)\times \int_{\vq_1,\vq_2} P_{11}(q_1)P_{11}(q_2)/q^2\,,\nn\\
  B_{521}^{I,hh} &\sim& k^4P_{11}(k_2)\times \int_{\vq_1,\vq_2} P_{11}(q_1)[P_{11}(q_2)]^2/q^4\,,\nn\\
  B_{521}^{II,hh} &\sim& k^2P_{11}(k_2)P_{11}(k_3)\times \int_{\vq_1,\vq_2} P_{11}(q_1)P_{11}(q_2)/q^2\,,\nn\\
  B_{431}^{I,hh} &\sim& k^4P_{11}(k_2)\times \int_{\vq_1,\vq_2} P_{11}(q_1)[P_{11}(q_2)]^2/q^4\,,\nn\\
  B_{431}^{II,hh} &\sim& k^4P_{11}(k_2)P_{11}(k_3)\times \int_{\vq_1} P_{11}(q_1)/q_1^2 \times \int_{\vq_2} P_{11}(q_2)/q_2^2\,,\nn\\
  B_{431}^{III,hh} &\sim& k^4P_{11}(k_2)\times \int_{\vq_1,\vq_2} P_{11}(q_1)P_{11}(q_2)P_{11}(|\vq_1+\vq_2|)/q^4\,,\nn\\
  B_{332}^{I,hh} &\sim& k^6\times \int_{\vq_1,\vq_2} [P_{11}(q_1)]^2P_{11}(q_2)P_{11}(|\vq_1+\vq_2|)/q^6\,,\nn\\
  B_{332}^{II,hh} &\sim& k^4P_{11}(k_2)P_{11}(k_3)\times \int_{\vq_1} P_{11}(q_1)/q_1^2 \times \int_{\vq_2} P_{11}(q_2)/q_2^2\,,\nn\\
  B_{332}^{III,hh} &\sim& k^6P_{11}(k_2)\times \int_{\vq_1} P_{11}(q_1)/q_1^2 \times \int_{\vq_2} [P_{11}(q_2)]^2/q_2^4\,,\nn\\
  B_{422}^{I,hh} &\sim& k^6\times \int_{\vq_1,\vq_2} P_{11}(q_1)[P_{11}(q_2)]^3/q^6\,,\nn\\
  B_{422}^{II,hh} &\sim& k^6\times \int_{\vq_1,\vq_2} [P_{11}(q_1)]^2[P_{11}(q_2)]^2/q^6\,.
  \eea
The leading UV dependence arises from $B_{611}$ and $B_{521}^{II}$, carrying a single factor of $k^2/q^2$, analogously to $P_{15}$ for the power spectrum. Indeed, $B_{521}^{II}$ is exactly proportional to $P_{15}(k_2)$, while for $B_{611}$ only the scaling is identical, but not the precise form. These leading terms will be discussed in detail next. Before that, we briefly comment on the contributions with subleading UV dependence, and their EFT counterpart. The terms with the smallest UV dependence are those where all input power spectra $P_{11}$ are evaluated for wavenumbers that become large in the double-hard limit, i.e.\ for the loop wavevectors $\vq_1$, $\vq_2$ or some linear combination of them. There are three contributions of this type, being $B_{422}^I$, $B_{422}^{II}$ and $B_{332}^I$. They carry a $k^6/q^6$ suppression factor, and would correspond to pure noise terms in the EFT setup. $B_{332}^{III}$ is also proportional to $k^6$, but arises from a product of a one-loop $P_{13}$ propagator term as well as a one-loop noise term. Nevertheless, due to the $k^6$ scaling, it is extremely suppressed in the double-hard limit. Finally, among the remaining terms, that are all involving a $k^4$ factor, one can discriminate two categories: $B_{521}^I$, $B_{431}^I$ and $B_{431}^{III}$ correspond to cross terms between stochastic noise and propagator EFT corrections, which may be considered as a stochastic contribution to $\cssq$~\cite{Cabass:2019lqx}, while $B_{431}^{II}$ and $B_{332}^{II}$ have the form of a ``propagator correction squared''. One may expect these last contributions to feature the largest UV sensitivity apart from the leading terms $B_{611}$ and $B_{521}^{II}$. In the following we consider EFT corrections only for the leading UV sensitive contributions, $B_{611}$ and $B_{521}^{II}$, and comment on the impact of subleading terms in section\,\ref{sec:k4}. 

\begin{figure}
    \centering
    \includegraphics[width=0.99\textwidth]{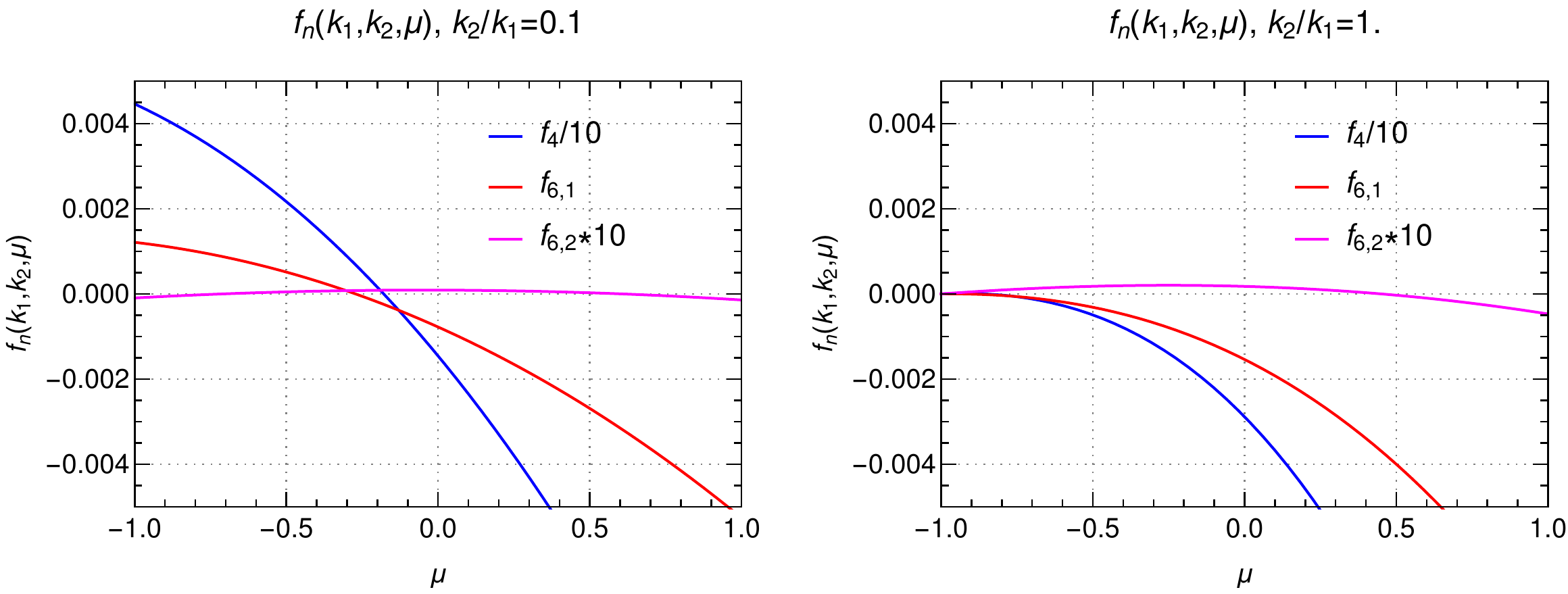}
    \caption{Shape functions $f_{6,1/2}({\bf k}_1,{\bf k}_2)$ related to the double-hard limit of $F^\text{(s)}_6({\bf k}_1,{\bf k}_2,{\bf q}_1,-{\bf q}_1,{\bf q}_2,-{\bf q}_2)$. We also show $f_4({\bf k}_1,{\bf k}_2)$ derived from the hard limit of $F^\text{(s)}_4({\bf k}_1,{\bf k}_2,{\bf q},-{\bf q})$.
    Note that $f_4$ is divided by a factor $10$, and $f_{6,2}$ multiplied by $10$ for better visibility.
    }
    \label{fig:fn}
\end{figure}

Let us now turn to the contributions with leading UV sensitivity at two-loop order. We start with $B_{611}$ which contains the $F_6$ kernel. Using the SPT recursion relations and a sequential Taylor expansion algorithm, we find that the double-hard limit 
 $q_1\propto q_2\to \infty$
 of the sixth order SPT kernel is given by
\bea\label{eq:F6hhlimit}
\int \frac{d\Omega_{q_1}}{4\pi}\frac{d\Omega_{q_2}}{4\pi}\,F^\text{(s)}_6(\vec k_1,\vec k_2,\vec q_1,-\vec q_1,\vec q_2,-\vec q_2) &\to& \frac{1}{q_1 q_2}\left(f_{6,1}(\vec k_1,\vec k_2)S_1(q_1/q_2)+f_{6,2}(\vec k_1,\vec k_2)S_2(q_1/q_2)\right)\,.
\eea
To our knowledge this analytical limit has not been presented in the literature yet.
The dependence on the external wavenumbers is encapsulated in the two shape functions $f_{6,1}(\vec k,\vec p)$
and $f_{6,2}(\vec k,\vec p)$, multiplying two functions $S_{1}(x)$ and $S_{2}(x)$ that we specify further below. The shape functions can be written as a linear combination of five basis functions $b^{(j)}(\vec k,\vec p)$, $j=1,\dots,5$, introduced in \eqref{eq:shapes},
\be
  f_i(\vec k,\vec p) = \sum_{j=1}^5 f_i^{(j)} \, b^{(j)}(\vec k,\vec p)\,,
\ee
or equivalently in the $E_{1,2,3}$ and $\Gamma$ basis.
Our analytical result for the coefficients for the two shape functions $f_{6,1}(\vec k,\vec p)$ and $f_{6,2}(\vec k,\vec p)$ are given in  Tab.\,\ref{tab:shape}.
As shown in section~\ref{sec:oneloop}, the function $f_4(\vec k,\vec p)$ appearing in the hard limit of the one-loop bispectrum can also be decomposed in the same way. The corresponding coefficients are also given in Tab.\,\ref{tab:shape}. In Fig.\,\ref{fig:fn} we show the dependence of $f_{6,1}(\vec k,\vec p)$ and $f_{6,2}(\vec k,\vec p)$ on $\mu$. It is apparent that these two shape functions are \emph{not} proportional to $f_4(\vec k,\vec p)$. As we will argue below, this implies that the UV-inspired approach cannot be extended to two-loop order. On the other hand, the  symmetry-based approach contains already all basis shapes $E_{1,2,3}$ and $\Gamma$ that are also required to absorb the double-hard limit of $B_{611}$.

\begin{table}[t]
\begin{center}
$
\begin{array}{c|ccccc}
j & 1 & 2 & 3 & 4 & 5 \\ \hline 
f_{6,1}^{(j)} & -\frac{1394259753263}{1811404542543750} & -\frac{70647110404331}{23548259053068750} & -\frac{11191}{38697750} & -\frac{9685830431171}{7849419684356250} & -\frac{10098786522983}{23548259053068750} \\
f_{6,2}^{(j)} & \frac{104211446312}{11774129526534375} & -\frac{78591466504}{3924709842178125} & 0 & -\frac{80969969032}{3924709842178125} & -\frac{41622522056}{11774129526534375} \\
f_3^{(j)} & 5/7& 1& 1/2& 2/7& 0\\
f_4^{(j)} & -\frac{4901}{339570} & -\frac{115739}{2037420} & -\frac{61}{7560} & -\frac{1592}{56595} & -\frac{12409}{1018710}
\end{array}
$

\bigskip

$
\begin{array}{c|cccc}
 & e_1 & e_2 & e_3 & \gamma \\ \hline 
f_{6,1} & -\frac{158092425677}{336403700758125} & -\frac{5853641823383}{47096518106137500} & -\frac{28205520799243}{23548259053068750} & -\frac{11191}{19348875}  \\
f_{6,2} & \frac{664042208}{336403700758125} & -\frac{20811261028}{11774129526534375} & -\frac{34169022472}{905702271271875} & 0\\
f_3 & -\frac{97}{231} &-\frac{12}{77} & -\frac{68}{77} & 1\\
f_4 & -\frac{1733}{436590} & -\frac{1457}{407484} & -\frac{6997}{339570} & -\frac{61}{3780} 
\end{array}
$
\end{center}
\caption{\label{tab:shape}
Expansion coefficients of the shape functions relevant for the double-hard limit of the two-loop bispectrum ($f_{6,i}$ for $B_{611}$ and $f_3$ for $B_{521}$), as well as, for comparison, for the hard limit at one-loop ($f_4$ for $B_{411}$ and $f_3$ for $B_{321}$). The five basis functions are defined in \eqref{eq:shapes}. The lower table shows the expansion coefficients in the equivalent basis of $E_{1,2,3}$ and $\Gamma$.
}
\end{table}

Let us now return to the missing part in the discussion of the double-hard limit of the sixth order kernel in \eqref{eq:F6hhlimit}. The dependence on the ratio of the two ``hard'' loop wavenumbers $r=q_1/q_2$ is  given by the two functions
\bea\label{eq:S1S2}
  S_1(r) &=& -\frac{1}{716224 r^6}\Bigg[ 4 r (1 + r^2) (5760 + 13605 r^2 - 128258 r^4 + 13605 r^6 + 
     5760 r^8) \,, \nn\\
  && {} +   15 (r^2 - 1)^4 (384 + 2699 r^2 + 
     384 r^4) L(r) \Bigg] \,, \nn\\
  S_2(r) &=& -\frac{1}{ 512 r^6}\Bigg[4 r (1 + r^2) (105 - 340 r^2 + 406 r^4 - 340 r^6 + 105 r^8) \nn\\
  && {} + 15 (r^2 - 1)^4 (7 + 10 r^2 + 7 r^4) L(r)\Bigg] \,,
\eea
with $L(r)=\log((r-1)^2/(r+1)^2)$. 
Due to symmetry under exchange of $\vec q_1$ and $\vec q_2$,
the functions $S_1$ and $S_2$ satisfy
\be
  S_i(r) = S_i(1/r)\,.
\ee
The normalization is chosen such that $S_i(1)=1$. For $r\to 0$,
\be
  S_1(r)\to \frac{120424 r}{78337}+{\cal O}(r^3), \qquad S_2(r) \to \frac{64 r^3}{21}+{\cal O}(r^5)\,.
\ee
Note that $S_1$ appears also in the double-hard limit of $F_5^\text{(s)}$, known from the two-loop power spectrum \cite{Baldauf:2015aha}, 
\bea
  \int \frac{d\Omega_{q_1}}{4\pi}\frac{d\Omega_{q_2}}{4\pi}\,F^\text{(s)}_5(\vec k,\vec q_1,-\vec q_1,\vec q_2,-\vec q_2) &\to& -\frac{11191}{6449625}\frac{k^2}{q_1 q_2}S_1(q_1/q_2)\,.
\eea

The expansion of the SPT kernels can be used to obtain an analytical expression for the double-hard limit of the two-loop bispectrum $B_{2L}(k_1, k_2, k_3)$. Specifically, we are looking for contributions
that are parametrically suppressed by a factor $k^2$ compared to the tree-level bispectrum, in the limit $k_1\sim k_2\sim k_3\sim k\to 0$.
Only $B_{611}$ and $B_{521}$ yield such contributions, given by
\bea\label{eq:hhlimit}
  B_{611}^{hh}(k_1, k_2, k_3) &=& 90\left[f_{6,1}(\vec k_1,\vec k_2)s_1(\Lambda)+f_{6,2}(\vec k_1,\vec k_2)s_2(\Lambda)\right]P_{11}(k_1)P_{11}(k_2)\,, \nn\\
  B_{521}^{hh}(k_1, k_2, k_3) &=& 30\left[-\frac{11191}{6449625}(k_1^2+k_2^2)F^\text{(s)}_2(\vec k_1,\vec k_2)s_1(\Lambda)\right]P_{11}(k_1)P_{11}(k_2) \,,
\eea
where $F^\text{(s)}_2(\vec k_1,\vec k_2)$ denotes the usual SPT kernel. The double-hard limits  $B_{611}^{hh,s}$ and $B_{521}^{hh,s}$ are obtained by adding two permutations in each case\footnote{Note that we multiplied $B_{521}^{hh}$ by a factor two compared to $B_{521}$ from \eqref{eq:B2Lexpressions}, accounting for the trivial permutation of $\vk_1$ and $\vk_2$ in that limit. The symmetrized bispectrum is thus obtained by adding two permutations to $B_{521}^{hh}$, and five permutations to $B_{521}$, respectively.}.
Furthermore, note that $B_{521}^{hh}$ contains the shape function 
\be
  f_3(\vec k,\vec p)= (k^2+p^2)F^\text{(s)}_2(\vec k,\vec p)\,,
\ee
appearing also in $B_{321}^{h}$ at one-loop. As noted above, it can also be decomposed into the basis functions,
with coefficients given in the third line of Tab.\,\ref{tab:shape} for the $b^{(j)}$ or $E_i$ basis, respectively.

The coefficients $s_i(\Lambda)$ for $i=1,2$ denote the UV-sensitive two-loop integrals
\be
  s_i(\Lambda) = \int_{q_1,q_2<\Lambda} d^3q_1 d^3q_2 \frac{S_i(q_1/q_2)}{q_1 q_2} P_{11}(q_1) P_{11}(q_2)\,,
\ee
and encapsulate the dependence on the UV cutoff $\Lambda$. They can easily be evaluated numerically, using the result for $S_i(r)$. 
Parametrically, the scaling with the cutoff can be roughly estimated using the symmetry in $q_1\leftrightarrow q_2$ and the small/large-r limit of $S_i$,
\bea
  s_1(\Lambda) &\sim& \int_0^\Lambda dq_1\, P_{11}(q_1) \int_0^{q_1} dq_2 q_2^2 P_{11}(q_2) \sim \int_0^\Lambda dq\, P_{11}(q) \log^4(q/q_0)\,,\nn\\
  s_2(\Lambda) &\sim& \int_0^\Lambda dq_1 \frac{P_{11}(q_1)}{q_1^2} \int_0^{q_1} dq_2 q_2^4 P_{11}(q_2) \sim \int_0^\Lambda dq\, P_{11}(q) \log^3(q/q_0)\,,
\eea
where the latter estimate applies to a $\Lambda$CDM linear input power spectrum with asymptotic behaviour $P_{11}(q)\to q^{-3}\log^3(q/q_0)$.
Therefore, $s_i$ are more sensitive to the UV cutoff compared to the corresponding integral
$\sigmadsq\sim \int_0^\Lambda dq\, P_{11}(q)$ at one-loop, as for the power spectrum.

For the fiducial $\Lambda$CDM model considered in this work one finds 
\be\label{eq:s1s2num}
  s_1\simeq 52.51\, (\mbox{Mpc}/h)^2,\quad  s_2\simeq 29.55\, (\mbox{Mpc}/h)^2\,,
\ee
 for $\Lambda=0.6\ihMpc$.

\bigskip

The double-hard limit can be summarized as
\be
  B_{2L}(k_1, k_2, k_3) \to B_{2L,1}^{hh}(k_1, k_2, k_3)s_1(\Lambda) + B_{2L,2}^{hh}(k_1, k_2, k_3)s_2(\Lambda)+{\cal O}(k^4P_L(k)^2,k^4P_L(k),k^6)\,,
\ee
where, for $i=1,2$,
\be\label{B2Lhh}
  B_{2L,i}^{hh}(k_1, k_2, k_3) = b_{2L,i}^{hh}(\vec k_1,\vec k_2)P_{11}(k_1)P_{11}(k_2) + 2\,{\rm permutations}\,,
\ee
with
\bea\label{eq:b2Li}
  b_{2L,1}^{hh}(\vec k,\vec p) &=& 90f_{6,1}(\vec k,\vec p)-\frac{11191}{6449625}30(k^2+p^2)F^\text{(s)}_2(\vec k,\vec p)\,,\nn\\
  b_{2L,2}^{hh}(\vec k,\vec p) &=& 90 f_{6,2}(\vec k,\vec p)\,.
\eea
Both of these shape functions can be written as linear combinations of the five basis functions \eqref{eq:shapes}, or alternatively in the basis of $E_{1,2,3}$ and $\Gamma$.
The coefficients are given in  Tab\,\ref{tab:shape2}.

\begin{table}[t]
\begin{center}

$
\begin{array}{c|ccccc}
j & 1 & 2 & 3 & 4 & 5 \\ \hline 
b_{2L,1}^{hh} & -\frac{27853833395669}{261647322811875} & -\frac{84266949648881}{261647322811875} &
-\frac{22382}{429975} & -\frac{3660985992757}{29071924756875} &
-\frac{10098786522983}{261647322811875} 
\\
    & -0.10646 & -0.32206 & -0.052054 & -0.12593 & -0.038597\\ \hline
b_{2L,2}^{hh} & \frac{208422892624}{261647322811875} & -\frac{157182933008}{87215774270625} & 0 &
-\frac{161939938064}{87215774270625} & -\frac{83245044112}{261647322811875}\\
    & 0.00079658 & -0.0018022 & 0 & -0.0018568 & -0.00031816\\ \hline
b_{1L}^{h} & -\frac{52891}{56595} & -\frac{148618}{56595} & -\frac{61}{105} & -\frac{66706}{56595} &
-\frac{24818}{56595}\\
    & -0.93455 & -2.6260 & -0.58095 & -1.1787 & -0.43852
\end{array}
$

\bigskip

$
\begin{array}{c|cccc}
 & e_1 & e_2 & e_3 & \gamma \\ \hline 
b_{2L,1}^{hh} & -\frac{152780472044}{7475637794625} & -\frac{146227011253}{
  47572240511250} & -\frac{1244431601311}{
  20126717139375} & -\frac{44764}{429975} \\
   &  -0.020437 & -0.0030738 & -0.061830 & -0.10411 \\ \hline
b_{2L,2}^{hh} & \frac{1328084416}{7475637794625} & -\frac{41622522056}{
  261647322811875} & -\frac{68338044944}{20126717139375}&0\\
   & 0.00017766 & -0.00015908 & -0.0033954 & 0 \\
    \hline
b_{1L}^{h} & \frac{817}{8085} & -\frac{2161}{56595} & -\frac{12946}{  56595} & -\frac{122}{105} \\
    & 0.10105 & -0.038184 & -0.22875 & -1.1619
\end{array}
$
\end{center}
\caption{\label{tab:shape2}
Expansion coefficients of the two independent shape functions \eqref{eq:b2Li} for the double-hard limit of the complete two-loop bispectrum, as well as for the hard limit of the one-loop bispectrum. The coefficients in the upper table refer to the basis given in \eqref{eq:shapes}, and the one in the lower table to  the basis of $E_{1,2,3}$ and $\Gamma$.
We also provide approximate numerical values rounded to 5 digits.
}
\end{table}

For a $\Lambda$CDM spectrum and cutoff $\Lambda\gtrsim 0.5\ihMpc$, the integrals $s_1(\Lambda)$ and $s_2(\Lambda)$ are of comparable
size. Since the coefficients $b_{2L,2}^{hh,(j)}$ are much smaller in magnitude than for the first shape, 
\be
  |b_{2L,2}^{hh,(j)}|\ll |b_{2L,1}^{hh,(j)}|,\quad j=1\dots 5\,,
\ee
the contribution of the second shape is
suppressed compared to the first one, and contributes at most at the few percent level. We checked that, in practice, neglecting the second shape does not lead to any sizeable differences, but include it in our numerical analysis for completeness.

\bigskip

Let us now stress a difference in the renormalization of the two-loop power versus bispectrum.
For the two-loop power spectrum, the double-hard contributions are proportional to $k^2P_{11}(k)$, and can therefore be absorbed into the leading $\cssq$ EFT parameter.
For the two-loop bispectrum, the double-hard limit is given by a particular linear combination of either the basis functions $b^{(j)}$ defined in \eqref{eq:shapes}, or equivalently of the four shapes $E_{1,2,3}$ and $\Gamma$ that were introduced in the context of one-loop bispectrum renormalization.

However, the linear combination that appears for the two-loop bispectrum is different from the linear combination obtained from the UV limit of the one-loop bispectrum. While this may be expected in general, it is worthwhile to point out that this implies that the ``UV-inspired'' ansatz cannot be carried over to two loops. On the other hand, the ``symmetry-based'' approach is suited also to absorb the double-hard UV contributions at two-loop, since the coefficients of the $E_{1,2,3}$ and $\Gamma$ are treated as independent free parameters.

\begin{figure}
    \centering
    \includegraphics[width=0.49\textwidth]{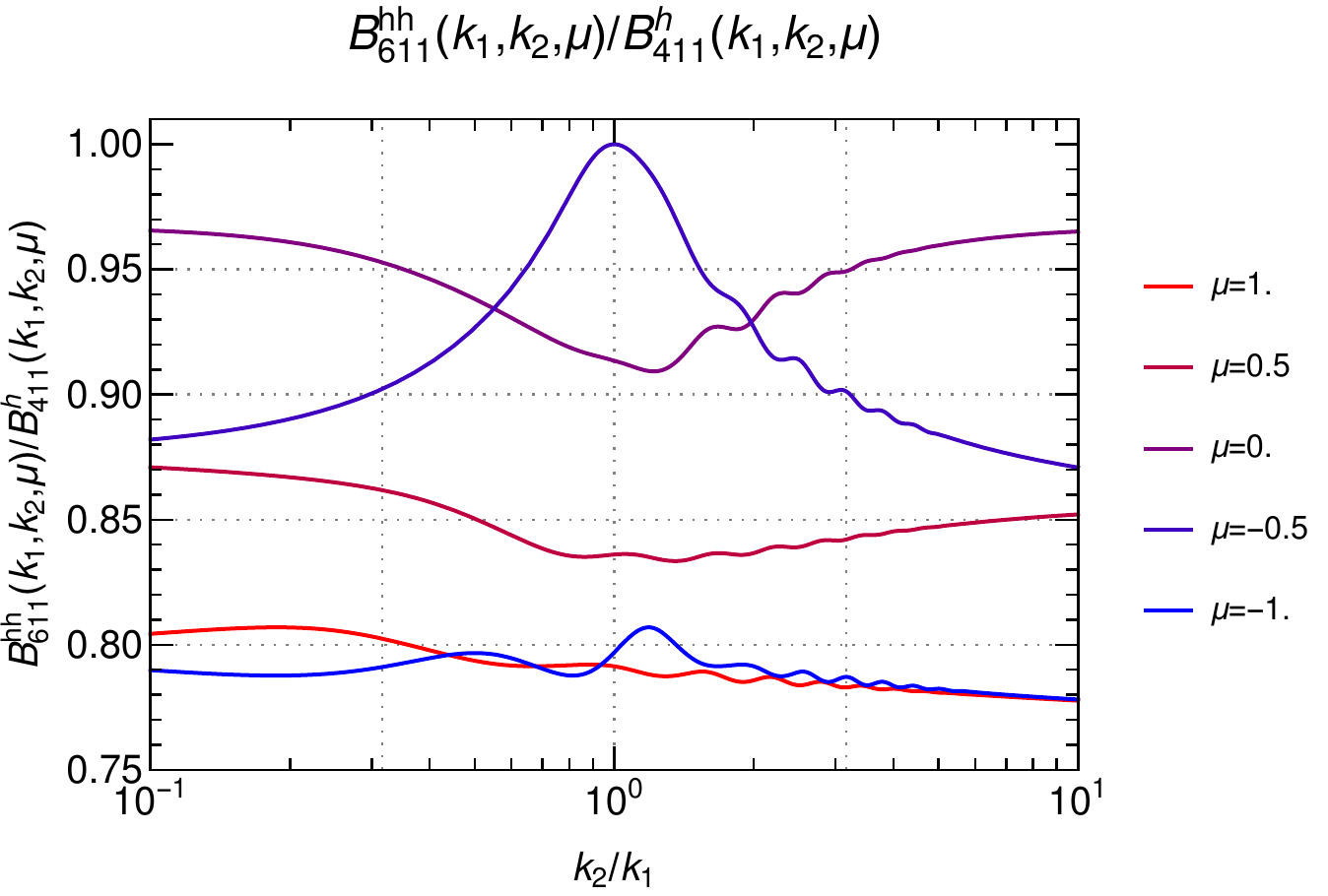}
    \includegraphics[width=0.49\textwidth]{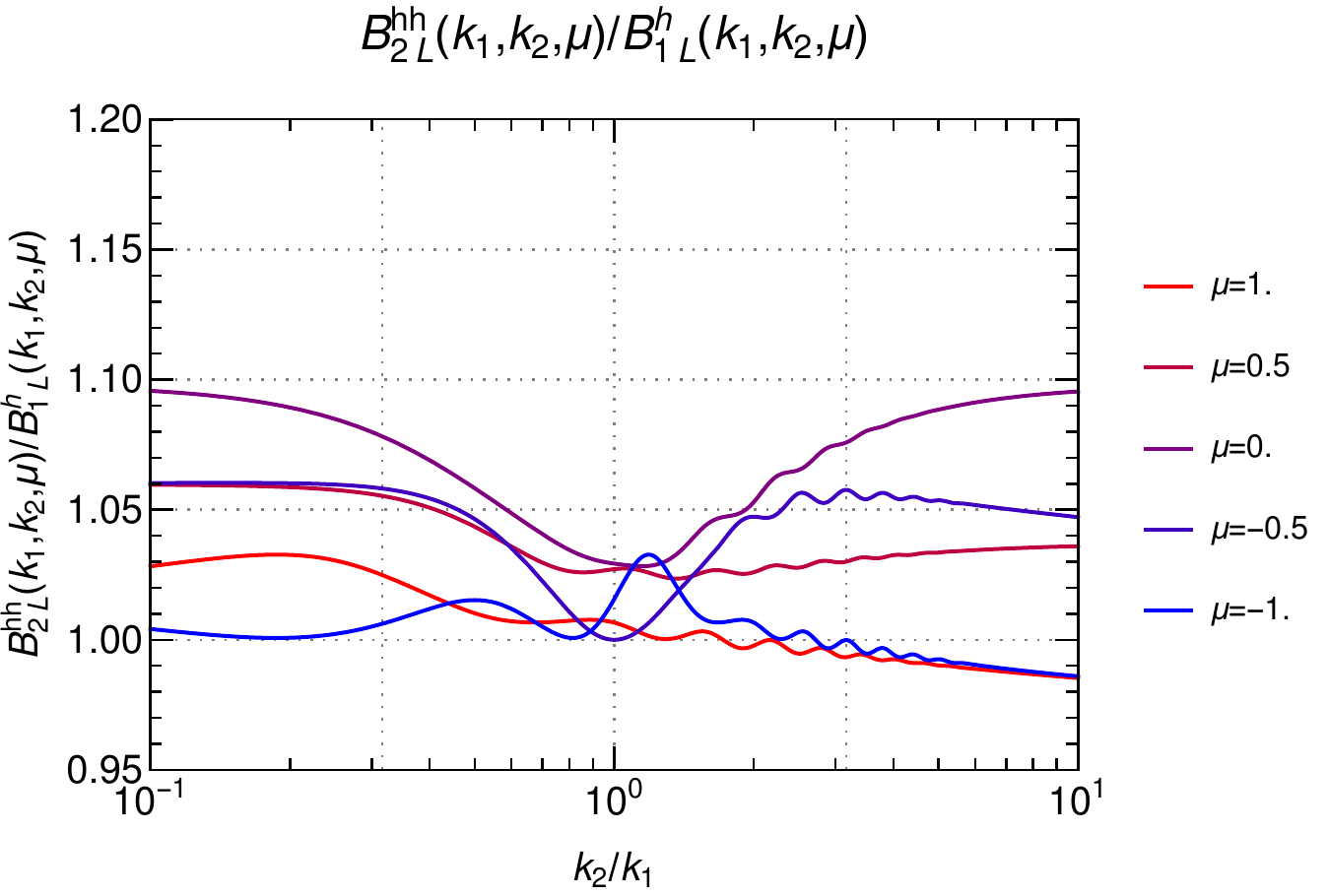}
    \\
    \includegraphics[width=0.49\textwidth]{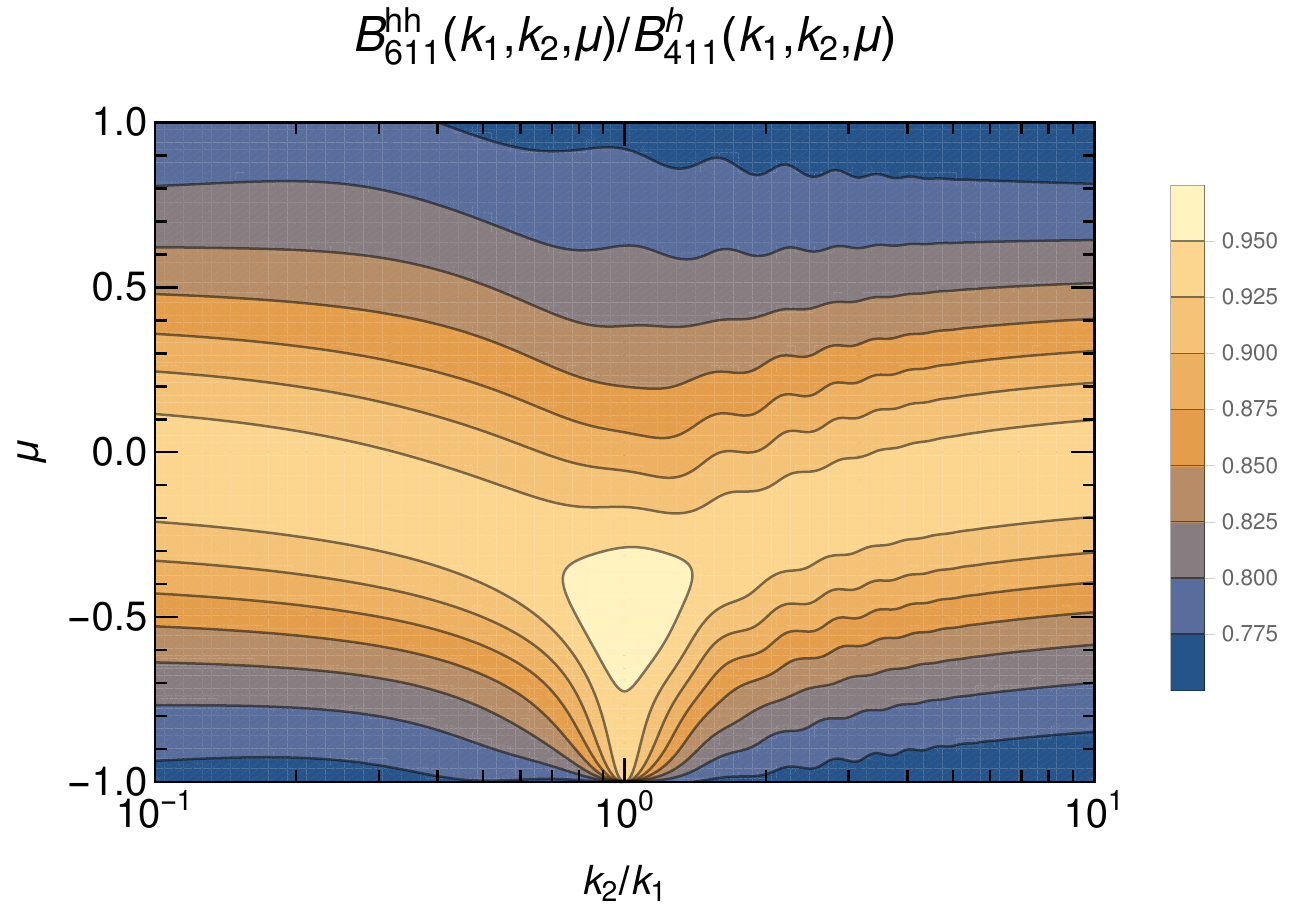}
    \includegraphics[width=0.49\textwidth]{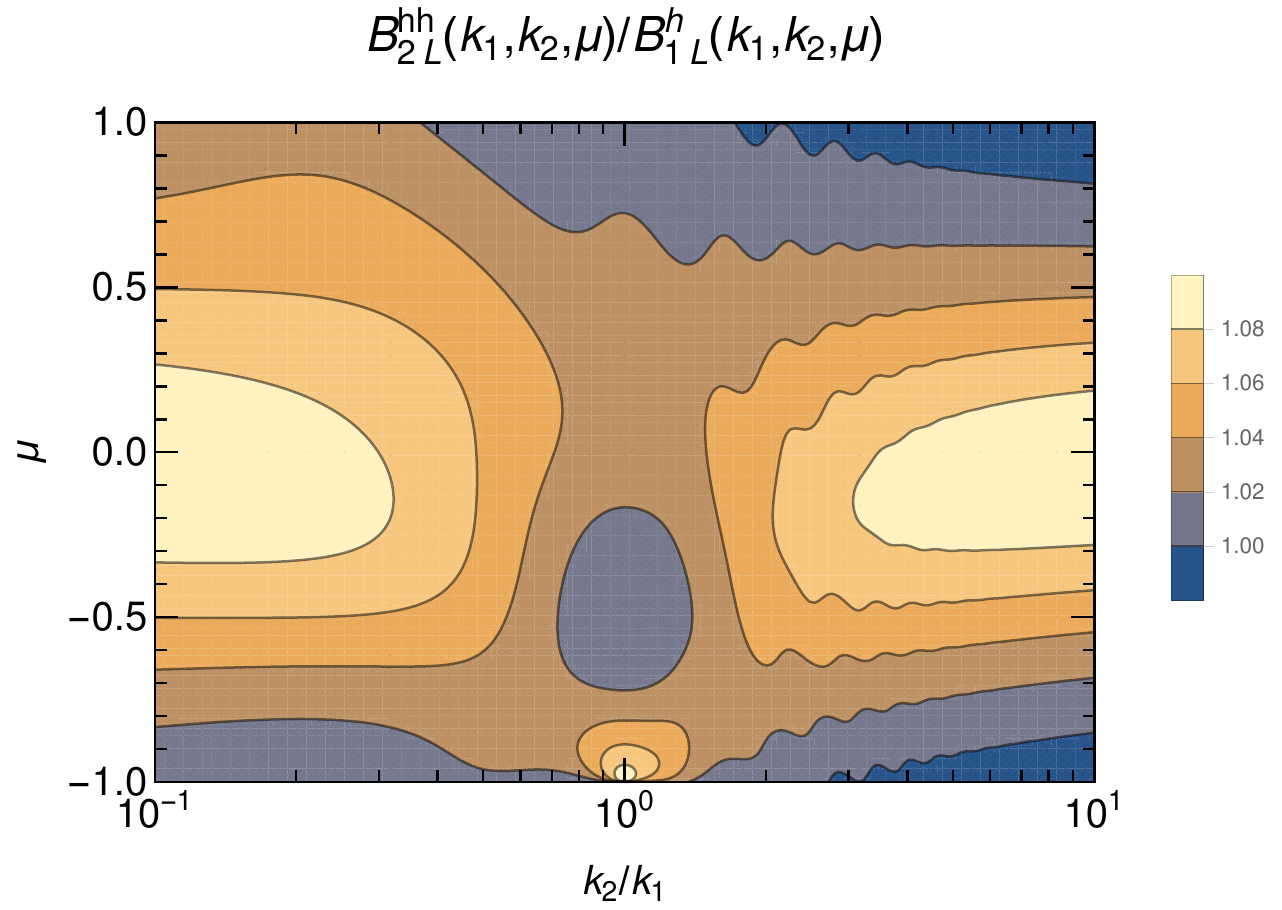}
   
    \caption{Double-hard limit of the two-loop bispectrum, relative to the single-hard limit of the one-loop bispectrum. We show the ratio of the symmetrized propagators $B_{611}^{hh,s}/B_{411}^{h,s}$ as well as the full bispectra $B_{2L}^{hh}/B_{1L}^h$. In all cases, $k_1=0.1\ihMpc$ and the ratio is normalized to the value in the equilateral configuration (i.e.\ for $k_2/k_1=1$ and $\mu=-0.5$).
    }
    \label{fig:doublehard}
\end{figure}

In order to illustrate by how much the shape-dependence of the double-hard limit at two-loop differs from the UV limit at one-loop, we show the ratio of the corresponding expressions in Fig.\,\ref{fig:doublehard}. Note that, in this figure, we include an arbitrary normalization factor (see legend), since we are interested only in the shape-dependence. We show the ratio both of the propagator-like contributions $B_{611}^{hh}/B_{411}^h$, as well as of the complete bispectrum, $B_{2L}^{hh,s}/B_{1L}^{h,s}$. The mismatch between the one- and two-loop hard limit is about up to $25\%$ in the former case, and slightly smaller, at the $\sim 10\%$ level, for the latter. To further quantify the mismatch of the hard limits, we include the expansion coefficients of $B_{1L}^h$ in Tab\,\ref{tab:shape2}. They are approximately, but not exactly, proportional to the ones of the dominant two-loop shape $B_{2L,1}^{hh}$.

In conclusion, we find that the two-loop bispectrum should be renormalized using the coefficients of the $E_{1,2,3}$ and $\Gamma$ shapes as independent, free parameters, corresponding to the ``symmetry-based'' approach. As we will discuss in the next section, completely renormalizing the two-loop bispectrum requires yet an additional parameter that accounts for the single-hard limit.

\subsection{Renormalization of the bispectrum}

Renormalization at two-loop requires to consider the double-hard (hh) as well as the single-hard (h) limit.
The hh contributions discussed above are proportional to a linear combination of the
shape functions \eqref{eq:shapes}, or equivalently the basis functions $E_{1,2,3}$ and $\Gamma$ known from the one-loop bispectrum renormalization. 

The symmetry-based one-loop renormalization involves four free EFT parameters related to these basis shapes.
The result for the hh limit implies that the two-loop bispectrum can be renormalized by the same four EFT parameters. 
For definiteness, we call these EFT parameters $c_i$. We can expand them as
\be
  c_i = c_i^{1L}+c_i^{2L}+\dots\,.
\ee 
Following \cite{Baldauf:2015aha}, we choose a renormalization condition such that $c_i^{2L}$ precisely cancel the hh
contribution to the bispectrum, while $c_i^{1L}$ are determined by fitting to simulations.
In practice, this means we consider the ``subtracted'' two-loop bispectrum, given by
\be\label{eq:B2Lsub}
  B_{2L}^{\text{sub}}(k_1, k_2, k_3;\Lambda) \equiv  B_{2L}(k_1, k_2, k_3;\Lambda) - B_{2L}^{hh}(k_1, k_2, k_3;\Lambda)\,,
\ee
where
\be
B_{2L}^{hh}(k_1, k_2, k_3;\Lambda) \equiv B_{2L,1}^{hh}(k_1, k_2, k_3)s_1(\Lambda) + B_{2L,2}^{hh}(k_1, k_2, k_3)s_2(\Lambda)\,,
\ee
with the RHS defined in \eqref{B2Lhh}. 

In addition, at two-loop, we need to consider the renormalization of the single-hard limit.
In principle, this should be possible by considering one-loop diagrams with an insertion of a one-loop
EFT operator. However, since an analytic treatment of the single-hard limit appears even more complex
as for the double-hard case, we restrict ourselves to a numerical evaluation, and follow the approach
proposed in \cite{Baldauf:2015aha}, that has been demonstrated to work well for the renormalization
of the two-loop power spectrum. 
To obtain the leading UV sensitive contributions in the single-hard limit, we consider the integrand of the
two-loop bispectrum, which can be written in the form
\be
  B_{2L}(k_1, k_2, k_3;\Lambda) = \int_{q_{1,2}<\Lambda} d^3q_1\,d^3q_2\,b_{2L}(k_1, k_2, k_3,\vec q_1,\vec q_2)P_{11}(q_1)P_{11}(q_2)\,.
\ee
Here $b_{2L}$ contains a sum of terms containing the SPT kernels as well as two further linear power spectra.
The leading UV dependence for $q_1\to\infty $ arises from those contributions containing a kernel of the form $F_n(\dots,\vec q_1,-\vec q_1,\dots)$,
which is suppressed as $1/q_1^2$ in that limit. All other contributions to $b_{2L}$ scale as $P_{11}(q_1)/q_1^4$. To obtain the leading UV dependence,
we therefore define
\be\label{eq:b2Lh}
  b_{2L}^h(k_1, k_2, k_3;\Lambda) \equiv {\cal N}\times \lim_{q_1\to\infty} q_1^2\int_{|q_{2}|<\Lambda} d^3q_2\,d\Omega_{q_1}\,b_{2L}(k_1, k_2, k_3,\vec q_1,\vec q_2) P_{11}(q_2)\,.
\ee
Here ${\cal N}$ is a normalization factor that we will specify shortly.
The asymptotic scaling of the kernels guarantees that the limit can be taken, and precisely accounts for the leading UV dependent contributions. 
In practice, we evaluate the RHS numerically, using the same algorithm as for the full two-loop bispectrum, except that
the magnitude of one of the loop wavenumbers is kept fixed at some large value (that is, much larger than $\Lambda$). We check that the choice for $q_1$ does not influence the result.
The single-hard limit of the two-loop bispectrum is then given by
\be\label{eq:B2Lhdef}
  B_{2L}^{h}(k_1, k_2, k_3;\Lambda) =  2\times\frac{3}{\cal N} \sigmadsq(\Lambda) b_{2L}^h(k_1, k_2, k_3;\Lambda)\,,
\ee
where $\sigmadsq(\Lambda)=\frac13\int_{q<\Lambda} d^3q P_{11}(q)/q^2$. The factor $2$ takes into account that either $q_1$ or $q_2$ can become large.
Following \cite{Baldauf:2015aha}, we assume that UV renormalization can be taken into account effectively by a shift in the value of $\sigmadsq$.
In particular, 
\be\label{eq:sigmadrep}
  \sigmadsq(\Lambda) \mapsto \sigmadsq(\Lambda) + \frac{210}{61}\gamma_{2-\text{loop}}(\Lambda)\,,
\ee
where $\gamma_{2-\text{loop}}(\Lambda)$ corresponds to the EFT contribution (with a conventional normalization factor related to the UV limit of the one-loop power spectrum).
Consequently, the renormalization of the single-hard limit yields an additional contribution to the bispectrum, of the form
\be
2\times\frac{3}{\cal N} \times \frac{210}{61}\gamma_{2-\text{loop}}(\Lambda) \, b_{2L}^h(k_1, k_2, k_3;\Lambda)\,.
\ee
This corresponds to the mixed contribution with an EFT parameter and a loop.
In principle, we could add this term to the two-loop result directly. However, one may notice that part of it is still degenerate with the double-hard contribution.
The reason is that for external wavenumbers far below $\Lambda$, the remaining loop integration in \eqref{eq:b2Lh} includes a hard region with $q_2\gg k_{1,2,3}$.
This region will yield contributions that are again proportional to the shape functions \eqref{eq:shapes} encountered in the double-hard limit.
According to the renormalization condition introduced above, these contributions must be cancelled by an appropriate choice of  $c_i^{2L}$.
In practice, this can be done by a subtraction analogous to that in \eqref{eq:B2Lsub}.
In particular, the contribution to \eqref{eq:b2Lh} in the limit $q_2\gg k_{1,2,3}$ can be obtained analytically from the double-hard limit considered before.
It is given by
\be\label{eq:b2Lhh}
b_{2L}^{hh}(k_1, k_2, k_3;\Lambda) = B_{2L,1}^{hh}(k_1, k_2, k_3)s_1^h(\Lambda) + B_{2L,2}^{hh}(k_1, k_2, k_3)s_2^h(\Lambda)\,,
\ee
where
\be\label{eq:sih}
  s_i^h(\Lambda) \equiv {\cal N}\times \lim_{q_1\to\infty} q_1^2\int_{q_{2}<\Lambda} d^3q_2\,d\Omega_{q_1}\, \frac{S_i(q_2/q_1)}{q_1 q_2} P_{11}(q_2)\,.
\ee
Using the analytical results \eqref{eq:S1S2}, the limit can be taken. 
One obtains
\bea
  s_1^h(\Lambda) &=& {\cal N}\times (4\pi) \times \frac{120424 }{78337} \, \int_{q_{2}<\Lambda} d^3q_2\, P_{11}(q_2)\,,\nn\\
  s_2^h(\Lambda) &=& {\cal N}\times (4\pi) \lim_{q_1\to\infty} q_1^2 \, \int_{q_{2}<\Lambda} d^3q_2 \frac{64}{21} \frac{q_2^2}{q_1^4} \, P_{11}(q_2)=0\,.
\eea
Thus, only the first line contributes in the limit $q_1\to\infty$. We note that, for the numerical evaluation, \eqref{eq:b2Lh} is computed in practice by fixing $q_1$
to some large but finite value, as mentioned above. Correspondingly, we compute $s_i^h(\Lambda)$ numerically using \eqref{eq:sih} with the same large, fixed value of $q_1$.
After all, we obtain the \emph{subtracted} contribution
\be\label{eq:b2Lhbar}
\bar b_{2L}^h(k_1, k_2, k_3;\Lambda) \equiv b_{2L}^h(k_1, k_2, k_3;\Lambda) - b_{2L}^{hh}(k_1, k_2, k_3;\Lambda)\,.
\ee
This finally gives the mixed one-loop/EFT contribution to the renormalized bispectrum,
\be\label{eq:B2Lctr}
  B_{2L}^{\text{ctr}}(k_1, k_2, k_3;\Lambda) \equiv  2\times\frac{3}{\cal N} \times \frac{210}{61} \gamma_{2-\text{loop}}(\Lambda) \, \bar b_{2L}^h(k_1, k_2, k_3;\Lambda)\,.
\ee
The normalization factor drops out when using the definition of $\bar b_{2L}^h$, and can be chosen by convenience.
In the following we adopt the choice ${\cal N}\equiv 6\times \frac{210}{61}$. 

\begin{figure}
\includegraphics[width=0.99\textwidth]{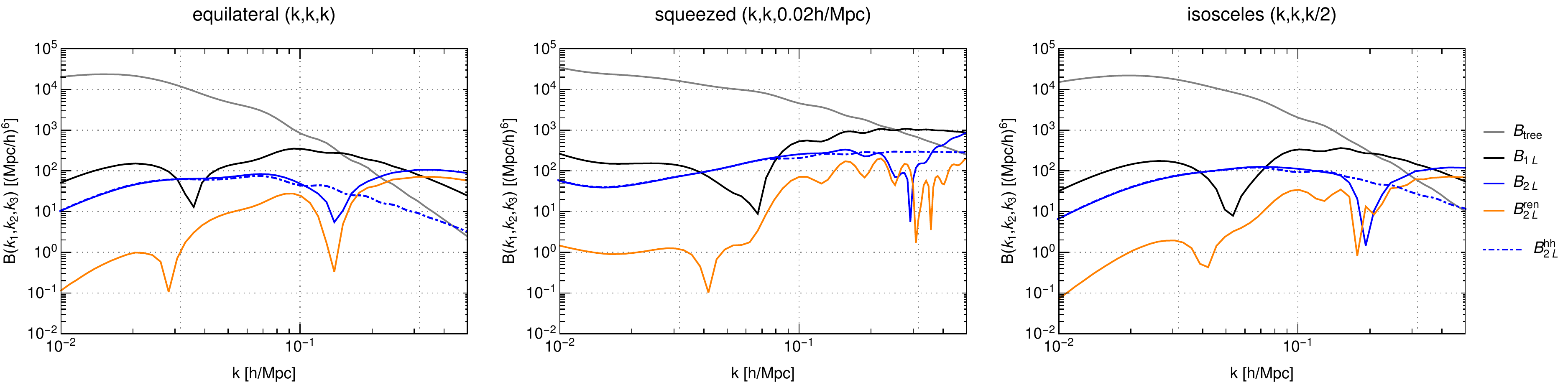}
\caption{\label{fig:2Lren} Renormalized two-loop contribution to the bispectrum $B_{2L}^{\text{ren}}$ (orange). For comparison we also show the unrenormalized SPT result (blue), as well as the one-loop (black) and tree-level (gray) contributions. The dot-dashed blue line shows the double-hard contributions $B_{2L}^{hh}$. Here we use $\bar \gamma_{2-\text{loop}}=1({\rm Mpc}/h)^2$ for $B_{2L}^{\text{ren}}$ for illustration.
}
\end{figure}

In summary, the total renormalized two-loop contribution to the bispectrum reads
\be
  B_{2L}^{\text{ren}}(k_1, k_2, k_3;\Lambda)=B_{2L}^{\text{sub}}(k_1, k_2, k_3;\Lambda)+B_{2L}^{\text{ctr}}(k_1, k_2, k_3;\Lambda)\,.
\ee
The only additional free parameter compared to the one-loop bispectrum is $\gamma_{2-\text{loop}}$. 
For illustration, we show the renormalized two-loop contribution to the power spectrum in Fig.\,\ref{fig:2Lren}, and compare it to the SPT tree-, one- and two-loop results. Note that the subtraction of the double-hard contribution almost completely cancels with the bare bispectrum for wavenumbers below about $0.1\ihMpc$, indicating that the SPT result within this regime is dominated by UV sensitive contributions.

\subsection{Cutoff independence}

The renormalized bispectrum should be independent of the cutoff $\Lambda$,
\be\label{eq:nocutoff}
  0 = \frac{d}{d\Lambda}B_{2L}^{\text{ren}}\,.
\ee
Following the usual renormalization procedure, the explicit dependence due to the cutoff in the loop integration should be cancelled by the
EFT parameters. The condition that the total result is cutoff independent then leads to (Wilsonian) renormalization group equations for the
EFT parameters $c_i$. The parameters $c_i\in\{\epsilon_i,\gamma_{1-\text{loop}}\}$ can be thought of as being split into a one- and two-loop part. The former accounts for the cutoff-dependence of the one-loop contribution, and corresponds to the numerical values for these parameters as quoted below. The latter is chosen implicitly to cancel the double-hard two-loop contributions within the renormalization scheme adopted here, and not needed explicitly. After the subtraction of double hard contributions, the remaining cutoff-dependence at two-loop order is related to the single hard limit only,
i.e.~\eqref{eq:nocutoff} leads to
\be
  0 = \frac{d}{d\Lambda}(\bar B_{2L}^{h}+B_{2L}^{\text{ctr}})=  \frac{d}{d\Lambda}\left(\frac{6}{\cal N} \sigmadsq(\Lambda) +\gamma_{2-\text{loop}}(\Lambda)\right)\bar b_{2L}^h(k_1, k_2, k_3;\Lambda)\,,
\ee
where $\bar B_{2L}^{h}$ is the bispectrum in the single-hard limit as defined in \eqref{eq:B2Lhdef}, but with $b_{2L}^h$ replaced by $\bar b_{2L}^h$, see \eqref{eq:b2Lhbar}, i.e. after subtracting remaining double-hard contributions. The latter are already accounted for within our renormalization scheme as explained above. Due to the subtraction, $\bar b_{2L}^h$ has formally no cutoff-dependence at leading power in gradients, to which we are working here (we shall confirm this numerically below). Consequently, we obtain the condition
\be
  0 = \frac{d}{d\Lambda}\left[\frac{6}{\cal N}\sigmadsq(\Lambda) + \gamma_{2-\text{loop}}(\Lambda)\right]\,.
\ee
Inserting the normalization factor ${\cal N}$
this yields a renormalization group equation for $\gamma_{2-\text{loop}}$,
\be
  \frac{d \gamma_{2-\text{loop}}(\Lambda)}{d\Lambda} = -\frac{61}{210}\frac{4\pi}{3}P_{11}(\Lambda)\,.
\ee
Its solution reads
\be\label{eq:csLam}
  \gamma_{2-\text{loop}}(\Lambda) = \bar \gamma_{2-\text{loop}} +  \frac{61}{210}\frac{4\pi}{3}\int_{\Lambda}^\infty dq P_{11}(q)\,,
\ee
where $\bar \gamma_{2-\text{loop}}$ is the EFT parameter obtained for $\Lambda\to\infty$.

\begin{figure}
\includegraphics[width=0.99\textwidth]{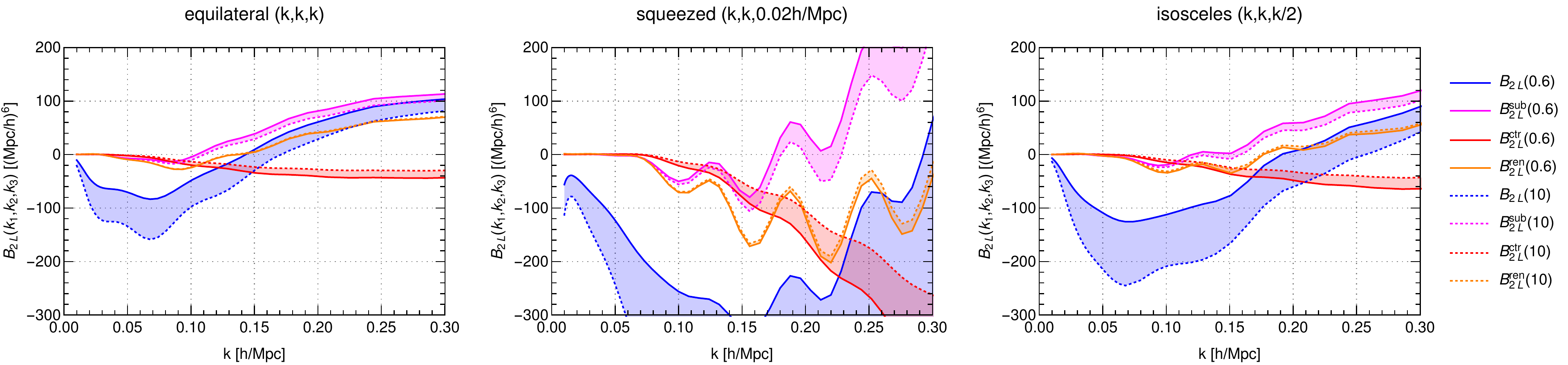}
\caption{\label{fig:cutoff} Two-loop bispectrum for UV cutoff $\Lambda=0.6\ihMpc$ (solid lines) and $\Lambda=10\ihMpc$ (dotted lines). The renormalized bispectrum $B_{2L}^{\text{ren}}=B_{2L}^{\text{sub}}+B_{2L}^{\text{ctr}}$ (orange) is approximately independent of the UV cutoff. Blue lines show the (bare) SPT two-loop result $B_{2L}$, which strongly depends on the UV cutoff. Magenta lines show the two-loop bispectrum after subtracting the double-hard contributions, $B_{2L}^{\text{sub}}=B_{2L}-B_{2L}^{hh}$. Red lines show $B_{2L}^{\text{ctr}}$ (for $\bar \gamma_{2-\text{loop}}=1({\rm Mpc}/h)^2$) which renormalizes remaining single-hard contributions.
}
\end{figure}

Since only the leading UV sensitive terms have been accounted for
in the renormalization procedure, it is an important check to which extent the dependence on the cutoff cancels in the numerical result. 
We find this to be the case to very good accuracy within the expected regime where all wavenumbers are far below the cutoff, $k_i\ll \Lambda$, see Fig.\,\ref{fig:cutoff}.
In particular, the dominant cutoff dependence of the (bare) SPT result (blue lines) is already removed when considering the subtracted bispectrum $B_{2L}^{\text{sub}}$ (magenta lines).
Nevertheless, $B_{2L}^{\text{sub}}$ shows a residual cutoff dependence (magenta dotted versus solid lines). Only when adding $B_{2L}^{\text{ctr}}$, with an EFT parameter $\gamma_{2-\text{loop}}(\Lambda)$ according to \eqref{eq:csLam}, the dependence on $\Lambda$ drops out in the renormalized bispectrum (solid versus dotted orange lines). Note that this cancellation
occurs for any choice of the free parameter $\bar \gamma_{2-\text{loop}}$ on the RHS of \eqref{eq:csLam}, since the ratio $B_{2L}^{\text{ctr}}/\gamma_{2-\text{loop}}(\Lambda)$)
is to a very good approximation independent of the cutoff. This observation is another consistency check, and in particular justifies adopting a cutoff-dependence of $\gamma_{2-\text{loop}}(\Lambda)$ as in \eqref{eq:csLam}.
The parameter $\bar \gamma_{2-\text{loop}}$ can be adjusted to reflect the actual impact of small-scale modes on the measurable large-scale bispectrum.
For a consistent interpretation, the renormalized bispectrum therefore has to be cutoff-independent for any value of $\bar \gamma_{2-\text{loop}}$. It is reassuring that this property is
indeed satisfied by the renormalization procedure adopted here.

\subsection{$k^4$ terms}\label{sec:k4}

In our discussion we restricted ourselves to the leading UV dependence, given by terms that are parametrically suppressed by the factor $k^2\sigmadsq$ relative to the lowest order for small external wavenumbers $k\propto k_i\to 0$. 

At one-loop, there are additional ``noise'' terms that scale as $k^4/P_{11}(k)\times \int d^3q P_{11}(q)^2/q^4$ relative to the lowest order. The UV sensitivity of the corresponding integral is extremely small for $\Lambda$CDM cosmologies, and therefore they are expected to be well captured by SPT. This means, while it would be possible to introduce EFT parameters related to stochastic noise terms in the effective stress tensor (that describe the generation of long-wavelength perturbations from mode-coupling interactions of a pair of small-scale modes with almost opposite wavenumber), this contribution is in practice negligible \cite{Baldauf:2015aha}.

\begin{figure}
    \centering
    \includegraphics[width=0.49\textwidth]{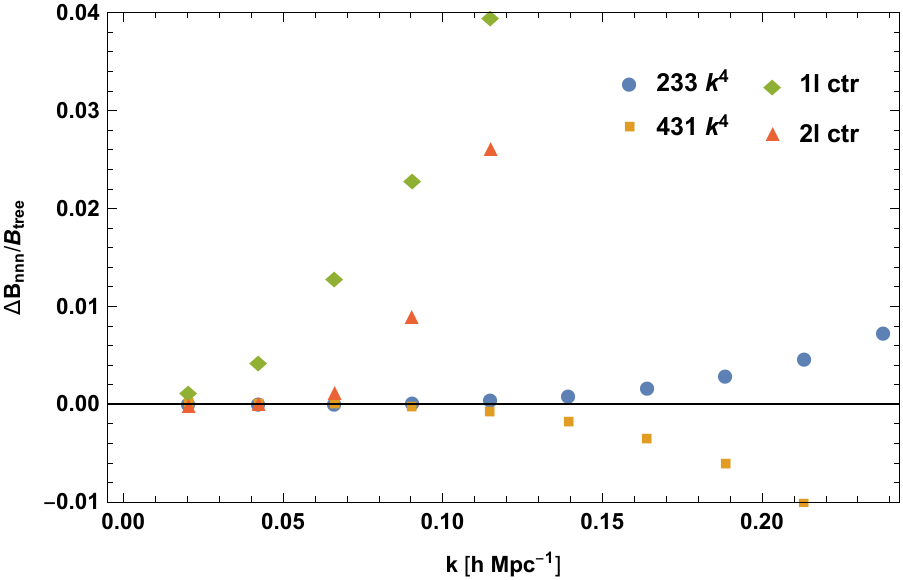}
    \caption{Size of the ``$(\cssq k^2)^2$'' contributions $B_{431}^{II}$ and $B_{332}^{II}$, compared to the one- and two-loop EFT contributions $B_{1L}^\text{ctr}$ and $B_{2L}^\text{ctr}$, normalized to the tree-level bispectrum, and evaluated for the equilateral shape at $z=0$.}
    \label{fig:k4terms}
\end{figure}

At two-loop, there are two contributions that scale as $k^4$, but are not related to noise terms, given by $B_{431}^{II}$ and $B_{332}^{II}$. These terms feature two loop integrals that are separated from each other, and can therefore be related to the square of two one-loop integrals. Each of them features a UV sensitivity typical of a one-loop propagator term, and therefore their hard limits can be treated on the same footing, however involving ``$(\cssq k^2)^2$'' terms. In particular,
\bea
  B_{431}^{II}(k_1,k_2,k_3) &=& B_{411}(k_1,k_2,k_3) \times P_{13}(k_2)/P_{11}(k_2)\,,\nn\\
  B_{332}^{II}(k_1,k_2,k_3) &=& B_{211}(k_1,k_2,k_3) \times [P_{13}(k_2)/P_{11}(k_2)]\times [P_{13}(k_3)/P_{11}(k_3)]\,,
\eea
where $P_{13}(k)/P_{11}(k)=3\int_{\vq} F_3^\text{(s)}(\vk,\vq,-\vq)P_{11}(q)$. The UV sensitivity arising from the single-hard limit of these expressions is already captured by the EFT contribution $B_{2L}^\text{ctr}$ introduced above. The double-hard limit is obtained by replacing $B_{411}\to B_{411}^h$ and $P_{13}\to P_{13}^h=-\frac{61}{210}k^2\sigmadsq$.
The corresponding EFT contributions can be estimated by replacing $\sigmadsq\mapsto\sigmadsq+ \frac{210}{61}\cssq$.

We have evaluated the double-hard contributions from the $B_{431}$ and $B_{322}$ diagrams and find that they are extremely small compared to the other EFT contributions for $k<0.15\ihMpc$, see Fig.\,\ref{fig:k4terms}.
Therefore, we do not include ``$(\cssq k^2)^2$'' terms in our numerical analysis.

\subsection{Summary}

In summary, the one- and two-loop bispectrum including leading EFT corrections are given by
\bea\label{eq:B2Lren}
  B_{1L}^{\text{ren}}(k_1,k_2,k_3;\gamma_{1-\text{loop}},\epsilon_i) &=& B_\text{tree}(k_1,k_2,k_3)+B_{1L}(k_1,k_2,k_3)+B_{1L}^{\text{ctr}}(k_1,k_2,k_3;\gamma_{1-\text{loop}},\epsilon_i)\,,\nn\\
  B_{2L}^{\text{ren}}(k_1,k_2,k_3;\gamma_{1/2-\text{loop}},\epsilon_i) &=& B_{1L}^{\text{ren}}(k_1,k_2,k_3;\gamma_{1-\text{loop}},\epsilon_i)\nn\\ 
  && {} +B_{2L}^{\text{sub}}(k_1,k_2,k_3)+B_{2L}^{\text{ctr}}(k_1,k_2,k_3;\gamma_{2-\text{loop}})\,,
\eea
where $B_{1L}^{\text{ctr}}$ depends on the EFT parameters $\epsilon_1,\epsilon_2,\epsilon_3$ and $\gamma_{1-\text{loop}}$, see \eqref{eq:1Lctr}.
The new two-loop terms are 
\be
  B_{2L}^{\text{sub}}=B_{2L}-B_{2L}^{hh}\,,
\ee
being the difference of the SPT two-loop result and the double-hard limit, see \eqref{eq:B2Lsub}, and $B_{2L}^{\text{ctr}}$ containing an additional EFT parameter $\gamma_{2-\text{loop}}$ taking care of the UV sensitivity in the single-hard limit, see \eqref{eq:B2Lctr}. The free parameters are therefore
\bea
  \{ \gamma_{1-\text{loop}},\epsilon_1,\epsilon_2,\epsilon_3 \} \qquad\qquad \text{1L\ symmetry-based approach,}\label{eq:1Lsymm} \\
  \{ \gamma_{1-\text{loop}},\epsilon_1,\epsilon_2,\epsilon_3, \gamma_{2-\text{loop}}  \} \qquad\qquad \text{2L\ symmetry-based approach} \label{eq:2Lsymm} \,.
\eea
Even though it is necessary to treat the EFT parameters as independent in order to be able to absorb the hard one-loop as well as single- and double-hard two-loop contributions into them, we also consider a \emph{naive} extension of the UV-inspired ansatz that has been proposed for the one-loop bispectrum~\cite{Steele:2020tak} for comparison. Namely, we assume that $\epsilon_1,\epsilon_2,\epsilon_3$ are related to $\gamma_{1-\text{loop}}$ in the same way as for the UV-inspired one-loop case, see\,\eqref{eq:epsUVinspired}. In addition, as discussed in section~\ref{sec:UVinsp}, the remaining EFT term may be linked to the $\cssq$ known already from the power spectrum. Altogether, we consider also the following UV-inspired cases:
\bea
 \label{eq:1LUV1par}\{ \gamma_{1-\text{loop}} \} &\qquad\qquad& \text{1L\ UV-inspired 1-parameter with } \epsilon_i/\gamma_{1-\text{loop}}\ \text{fixed}\,, \\
 \label{eq:1LUV0par}\emptyset &\qquad\qquad& \text{1L\ UV-inspired 0-parameter with}\ \epsilon_i/\gamma_{1-\text{loop}}\ \text{fixed and}\ \gamma_{1-\text{loop}}=\cssq|_{P_{1L}}\,,\\
 \label{eq:2LUV1par}  \{  \gamma_{1-\text{loop}} \} &\qquad\qquad& \text{2L\ UV-inspired 1-parameter with}\ \epsilon_i/\gamma_{1-\text{loop}}\ \text{fixed and}\ \gamma_{1-\text{loop}}=\gamma_{2-\text{loop}}\,, \\
 \label{eq:2LUV0par}\emptyset &\qquad\qquad& \text{2L\ UV-inspired 0-parameter with}\ \epsilon_i/\gamma_{1-\text{loop}}\ \text{fixed and}\  \gamma_{1-\text{loop}}=\gamma_{2-\text{loop}}=\cssq|_{P_{1L}}\,.
\eea
We emphasize that the EFT terms $\epsilon_1,\epsilon_2,\epsilon_3$ and $\gamma_{1-\text{loop}}$ account for the UV sensitivity of both the one-loop as well as the double-hard limit of the two-loop bispectrum.
The naive UV-inspired scheme therefore corresponds to the assumption that the leading UV sensitivity is generated by terms proportional to the hard limit $b_{1L}^h$ of the one-loop bispectrum, while hypothesizing that the shape functions $b_{2L,i}^{hh}$ corresponding to the double-hard limit do not contribute significantly (or at least the part of them that are not degenerate with $b_{1L}^h$). The UV-inspired approach is included only for illustration, since independence of the UV cutoff cannot in general be guaranteed within this approach.

We note that when using the symmetry-based approach, one could equivalently replace the subtracted by the full two-loop SPT bispectrum (i.e.\ omit the subtraction of $B_{2L}^{hh}$). From our analytical results presented above, we find that this would lead to a shift of the numerical values of the EFT parameters $\epsilon_1,\epsilon_2,\epsilon_3$ and $\gamma_{1-\text{loop}}$, but otherwise not change the result.

\section{Numerics}\label{sec:numerics}

\subsection{Setup}

We are using a realization-based calculation of the tree-level and one-loop contributions to the bispectrum to cancel cosmic variance and allow for an accurate fit of the EFT parameters from a modest simulation volume. For this purpose, the perturbative density fields are calculated order by order
in grid-PT \cite{Roth:2011test,Taruya:2018jtk,Taruya:2020qoy,Steele:2020tak}.

We consider a suite of 14 Gadget simulations, following the gravitational evolution of $1024^3$ particles in a cubic box of dimension $1500 \hMpc$. The non-linear matter density field is computed by assigning dark matter particles to a cubic lattice using the Cloud-in-Cell (CIC) assignment method and correcting for the CIC window in Fourier space.
We calculate the perturbative density fields up to sixth order from the Gaussian initial conditions that seeded the N-body simulations. To avoid aliasing, we cut off the linear density field at $\Lambda=0.3\ihMpc$ (but we also explore $\Lambda=0.6\ihMpc$ below). This wavenumber cutoff is implemented consistently, and the ability of the EFT to capture the cutoff dependence makes our final results cutoff independent. 

We bin the wavenumbers up to $0.3\ihMpc$ into ten linear bins and measure bispectra as a function of $k_1$, $k_2$ and $k_3$ using a FFT based estimator \cite{Baldauf:2014qfa,Steele:2020tak}. This bispectrum estimator is applied both to the non-linear density field as well as the perturbative density fields.

\begin{figure}
    \centering
    \includegraphics[width=0.32\textwidth]{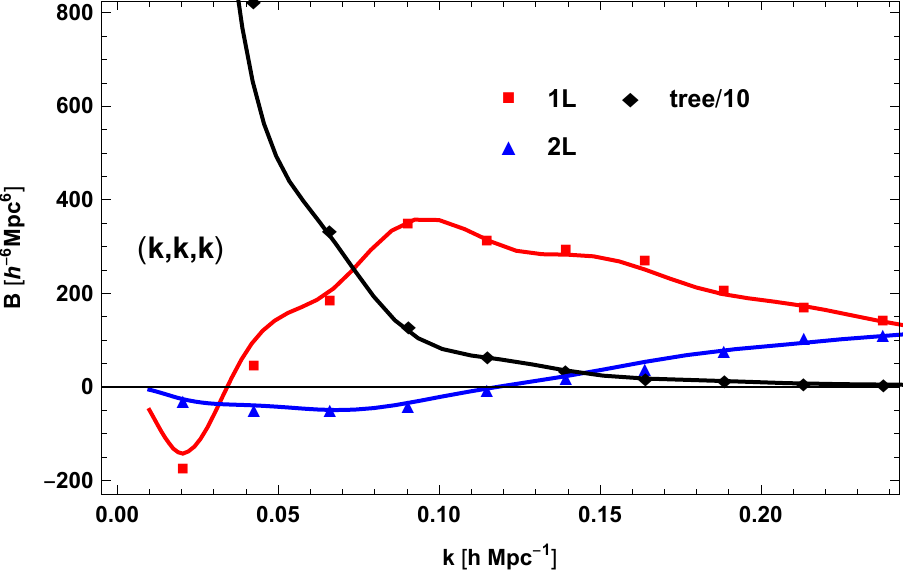}
    \includegraphics[width=0.32\textwidth]{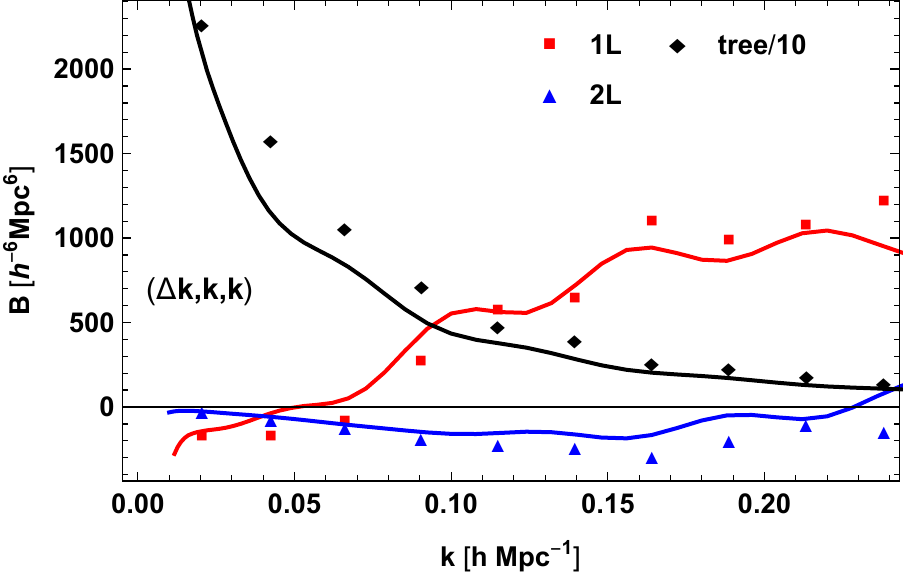}
    \includegraphics[width=0.32\textwidth]{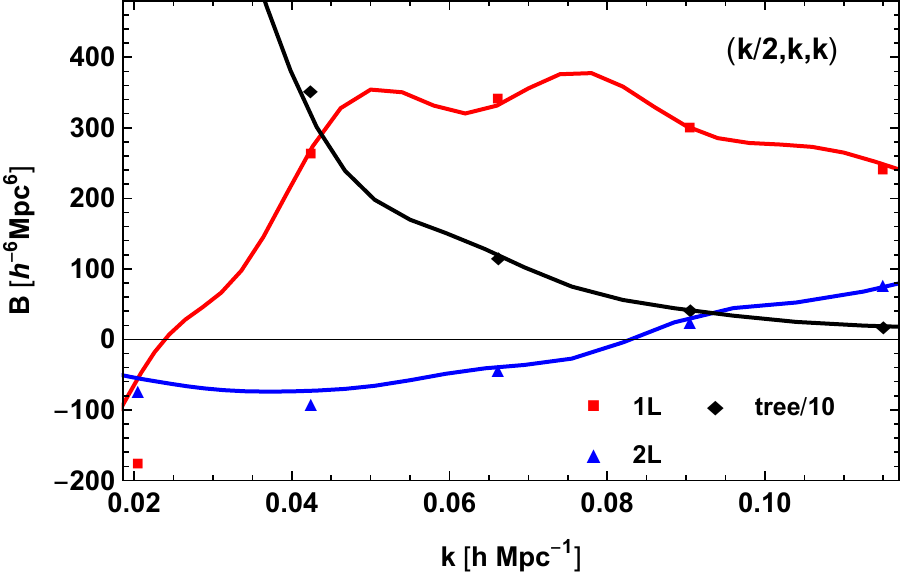}
    \caption{
    Comparison of the grid (symbols) and CUBA (lines) bispectrum calculations at one- and two-loops. The CUBA evaluations are based on effective wavenumbers whereas the grid calculation averages over bins in $k_1$, $k_2$ and $k_3$, leading to small differences between both approaches even at tree-level. The loop results are consistent with being identical up to bin averaging effects.
    }
    \label{fig:bnnntermssimcuba} 
\end{figure}

\begin{figure}
    \centering
    \includegraphics[width=0.6\textwidth]{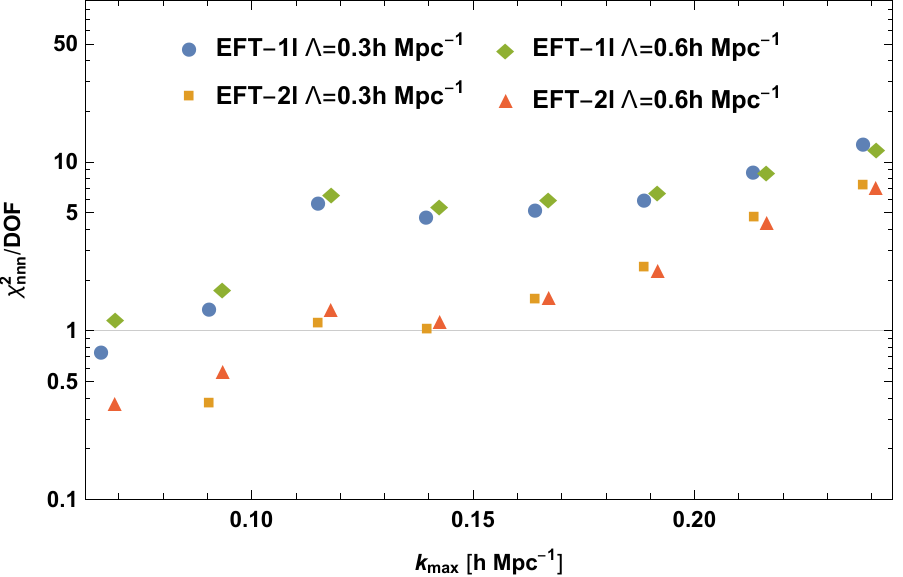}
    \caption{
    $\chi^2$ for the EFT one- and two-loop bispectra computed for a set of triangles $(k_1,k_2,k_3)$ with side length up to $k_\text{max}$, relative to N-body simulation results. The EFT terms account for the UV sensitivity of the SPT two-loop integrals, leading to consistent results when using a UV cutoff $\Lambda=0.3\ihMpc$ or $\Lambda=0.6\ihMpc$, respectively. 
    }
    \label{fig:chi2_cutoff}
\end{figure}

We have compared the CUBA evaluations of the one- and two-loop terms with the measurements on the grid and find good agreement, see Fig.\,\ref{fig:bnnntermssimcuba}.
As stated in \cite{Steele:2020tak}, there can be small time integration errors in the Gadget simulations which affect the subtraction of the leading perturbative contributions to the bispectrum. These errors are usually subdominant compared to cosmic variance, but do matter in the realization-based comparison used here. In addition to the EFT parameters we thus introduce growth factor corrections $\Delta D_i$ as free parameters in the SPT modelling 
\begin{equation}
    \delta_\text{n}(\vec x,t)=(1+\Delta D_1)\delta_1(\vec x,t)+(1+\Delta D_2)\delta_2(\vec x,t)+\delta_3(\vec x,t)+\delta_4(\vec x,t)~.
\end{equation}
We then determine the EFT parameters together with $\Delta D_i$ by minimizing
\begin{equation}
\begin{split}
\chi_{\mathrm{nnn}}^{2}(k_{\mathrm{max}})&=\sum_{k_{1,2,3}=k_{\mathrm{min}}}^{k_{\mathrm{max}}}\frac{1}{\Delta B^{2}_{\mathrm{nnn}}(k_{1},k_{2},k_{3})}\Bigl[B_{\mathrm{nnn}}(k_{1},k_{2},k_{3})-B_{\mathrm{SPT,reg}}(k_{1},k_{2},k_{3};\Delta D_{1},\Delta D_{2})\\
&~-B_\text{EFT}(k_{1},k_{2},k_{3};\gamma_{1/2-\text{loop}},\epsilon_i)\Bigr]^{2}\label{chinnn}~
\end{split}
\end{equation}
where $B_\text{nnn}$ stands for the N-body result and 
\bea
  B_{\text{SPT,reg}} &=& B_\text{tree}+B_{1L}+B_{2L}^{\text{sub}} \,,\nn\\
  B_\text{EFT} &=& B_{1L}^{\text{ctr}}(k_{1},k_{2},k_{3};\gamma_{1-\text{loop}},\epsilon_i)+B_{2L}^{\text{ctr}}(k_{1},k_{2},k_{3};\gamma_{2-\text{loop}})\,.
\eea
The growth factor corrections $\Delta D_i$ are taken into account in the tree-level and one-loop contribution following \cite{Steele:2020tak} (equation (59) therein).
The error $\Delta B_\text{nnn}$ is estimated from the variance of the bispectrum residual \cite{Steele:2020tak}.
In practice, we find a large degree of degeneracy among the $\epsilon_i$ contributions, and therefore fix $\epsilon_1=0$ even when following the symmetry-based approach. We checked that this restriction has only a minor impact on our results. 

\subsection{Results}

In Fig.\,\ref{fig:chi2_cutoff} we show the $\chi^2$ per degree of freedom (DOF) obtained from the one- and two-loop EFT bispectrum using the symmetry-based approach, see \eqref{eq:1Lsymm} and \eqref{eq:2Lsymm}, respectively, at $z=0$. We take our full set of configurations $(k_1,k_2,k_3)$ with $k_i\leq k_\text{max}$ into account, and show $\chi^2$/DOF versus $k_\text{max}$. The number of degrees of freedom is computed from the difference of the number of triangles contributing for a given $k_\text{max}$, and the number of free parameters, being four (five) for the one- and two-loop bispectrum, respectively, as well as two for the growth factor corrections $\Delta D_i$. 
For example for $k_\text{max}=0.1(0.2)\ihMpc$, our set comprises 65(369) triangles.

The residual is less than unity, indicating a $1\sigma$ agreement within the uncertainties, for $k_\text{max}\lesssim 0.08\ihMpc$ when adopting the one-loop approximation, consistent with earlier results~\cite{Steele:2020tak}. When adding the two-loop terms, the $1\sigma$ agreement extends up to wavenumbers of $k_\text{max}\lesssim 0.15\ihMpc$.

As an important check of the EFT approach, we repeated the analysis using a larger UV cutoff $\Lambda=0.6\ihMpc$ for comparison. Even through the SPT two-loop bispectrum shows a significant cutoff dependence, the EFT terms are able to absorb this UV sensitivity. This is demonstrated in Fig.\,\ref{fig:chi2_cutoff}  by a good agreement between the $\chi^2$ values obtained for two different choices of the UV cutoff.

\begin{figure}
    \centering
    \includegraphics[width=0.49\textwidth]{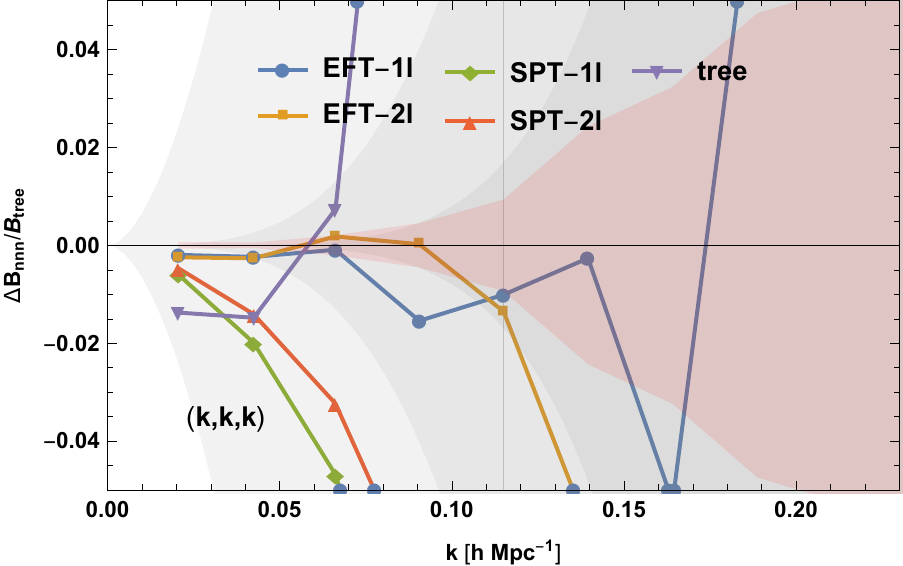}
    \includegraphics[width=0.49\textwidth]{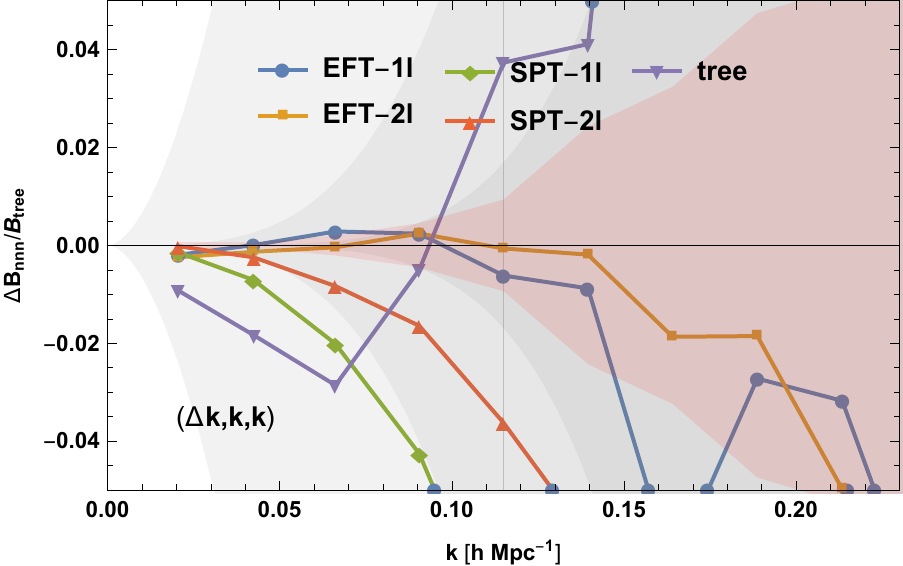}
    \includegraphics[width=0.49\textwidth]{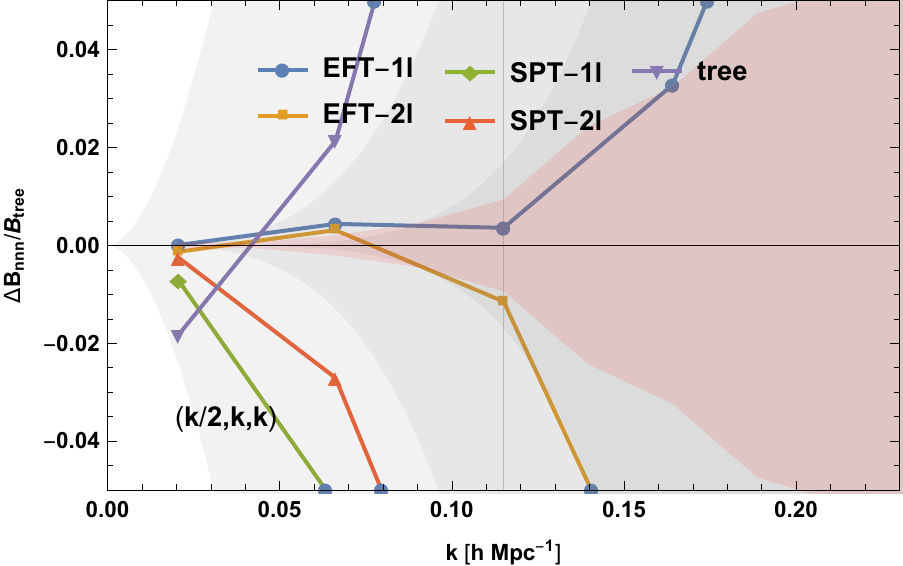}
    \caption{Difference between perturbative and N-body results, normalized to the tree-level bispectrum, for three different shapes: equilateral $(k,k,k)$, squeezed ($\Delta k,k,k$) with $\Delta k=0.02\ihMpc$, and isosceles $(k/2,k,k)$. The gray shaded region shows the expected theoretical uncertainty at tree-level, one- and two-loop, from light to dark gray, respectively~\cite{Baldauf:2016sjb}. The red shading indicates the error of the N-body simulation results. 
    }
    \label{fig:symmbased}
\end{figure}

The relative deviation of the SPT as well as EFT one- and two-loop bispectra from N-body simulation results is shown in Fig.\,\ref{fig:symmbased}, for three different shapes. For the EFT, the free parameters are kept fixed at the best-fit values obtained from the full set of triangles, for a pivot scale of $k_\text{pivot}=0.115\ihMpc$. As expected, apart from accounting for UV sensitivity, the EFT corrections extend the range of wavenumbers over which the perturbative result agrees well with N-body data. In addition, we checked that the remaining differences are consistent with the expected theoretical uncertainty due to missing higher-order corrections (viz. three-loop for the EFT two-loop result), as indicated by the gray shaded area~\cite{Baldauf:2016sjb}. Furthermore, we indicate the remaining variance in the N-body result by the red shaded area. The improvement of two- versus one-loop EFT is particularly relevant at relatively low wavenumbers, where the uncertainties are small. We note that also the SPT results tend to agree better with N-body data when including the two-loop piece, in particular for the squeezed configuration. Nevertheless, SPT deviates by more than percent level for wavenumbers around $0.05\ihMpc$, while the EFT two-loop result achieves this benchmark for wavenumbers below about $0.12-0.17\ihMpc$, depending on the configuration.

\begin{figure}
    \centering
    \includegraphics[width=0.49\textwidth]{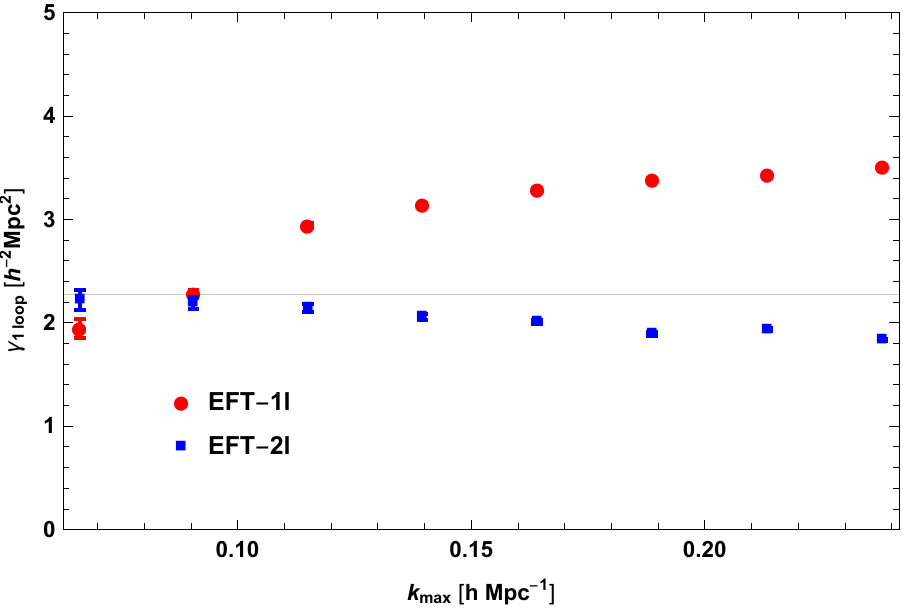}
    \includegraphics[width=0.49\textwidth]{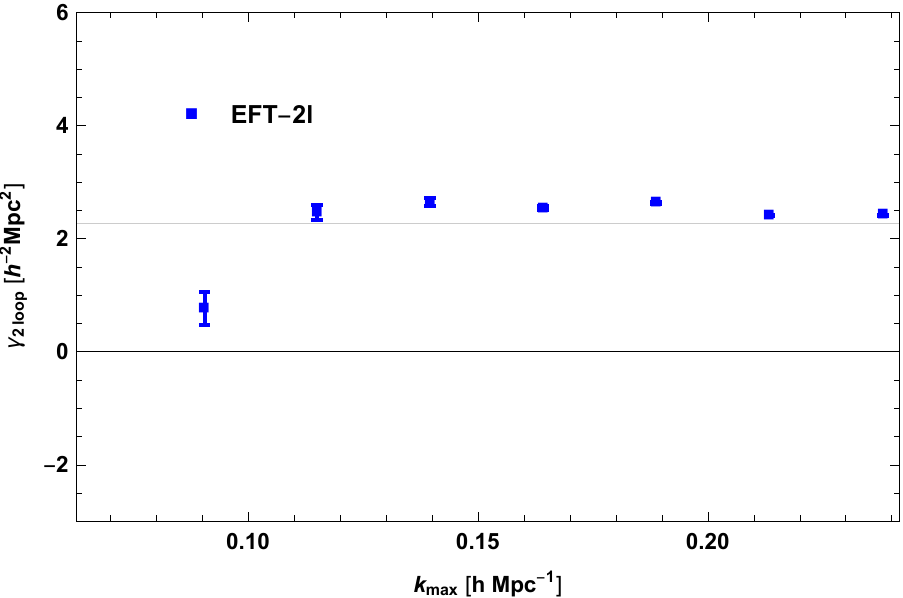}
    \includegraphics[width=0.49\textwidth]{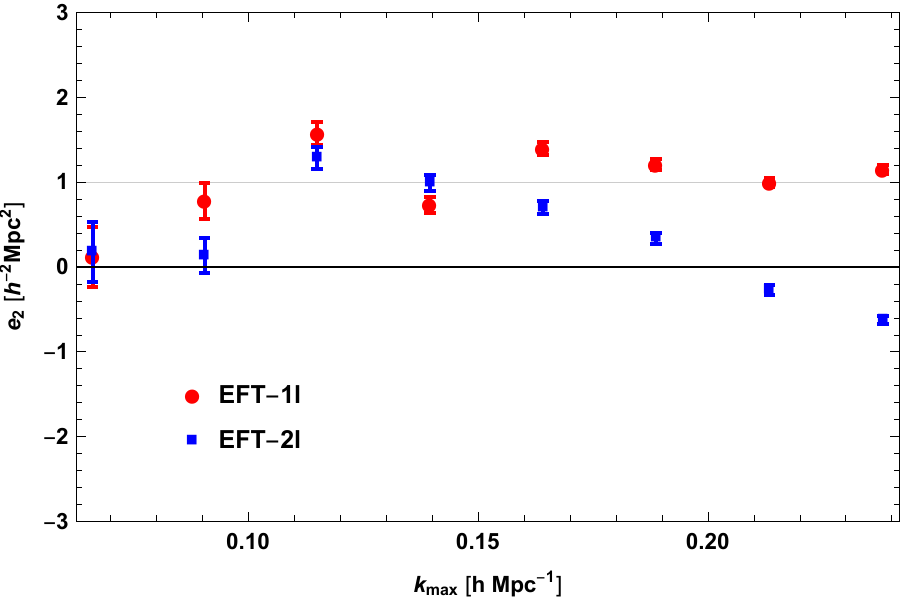}
    \includegraphics[width=0.49\textwidth]{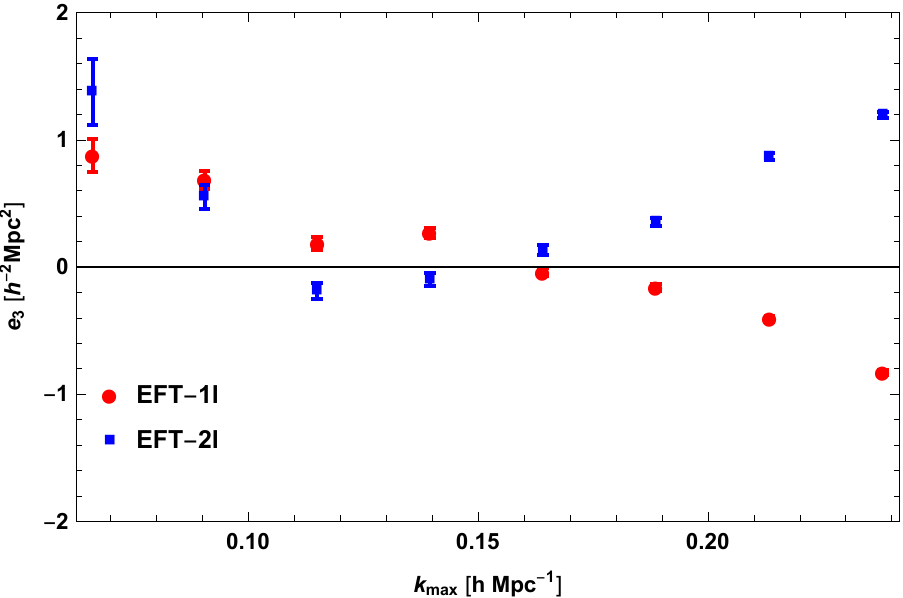}
    \caption{Best-fit values of the EFT parameters $\gamma_{1-\text{loop}}, \gamma_{2-\text{loop}}$, and $\epsilon_i$. The error bars indicate 1$\sigma$ uncertainties, and the gray shaded lines show the values to which the parameters would be fixed in the UV-inspired approach. Note that we fix $\epsilon_1=0$ to avoid numerical degeneracies. 
    }
    \label{fig:ctr}
\end{figure}

The best-fit values obtained for the EFT parameters are shown in Fig.\,\ref{fig:ctr}, as a function of $k_\text{max}$. The parameter $\gamma_{1-\text{loop}}$, that corresponds to the sound velocity correction at second order in perturbation theory, shows only a mild dependence over a rather wide range of values for $k_\text{max}$, especially at two-loop. This indicates that the two-loop terms indeed capture the shape-dependence well as compared to the N-body results, and allow for a reliable calibration of the EFT parameter $\gamma_{1-\text{loop}}$. In terms of the $\chi^2$ fit, this parameter also shows the smallest uncertainty, as indicated by the error bars in Fig.\,\ref{fig:ctr}.  

The EFT parameter $\gamma_{2-\text{loop}}$ is shown in the upper right panel of Fig.\,\ref{fig:ctr}. Its best-fit value is also rather stable over a wide range of $k_\text{max}$. However, for small values of $k_\text{max}$ it cannot be reliably determined, since $B_{2L}^\text{ctr}$ is extremely suppressed for small $k$. This is related to the subtraction of the overlap between single and double-hard contributions in~\eqref{eq:B2Lctr}, and the observation that the UV sensitivity of the two-loop bispectrum is dominated by the double-hard contributions in the limit $k_i\to 0$. In other words, for very small wavenumbers $B_{2L}^\text{ctr}(k_1,k_2,k_3)$ does not contribute to the two-loop bispectrum in a sizeable amount, and therefore a calibration of the corresponding EFT parameter $\gamma_{2-\text{loop}}$ requires to choose $k_\text{max}\gtrsim 0.12\ihMpc$.

The EFT parameters $\epsilon_2$ and $\epsilon_3$  are shown in the lower left and right panel of Fig.\,\ref{fig:ctr}, respectively. They are related to the operators $\Delta s^{ij}s_{ij}$ and $\partial_i[s^{ij}\partial_j\delta]$, involving the tidal tensor $s^{ij}$. For small $k_\text{max}$, they cannot be very precisely determined by the fit, as indicated by the error bars. This means in turn that they do not give a very large contribution for small wavenumbers. In addition, they are partially correlated with each other and with the $\gamma_{1/2-\text{loop}}$ parameters, making an interpretation of their best-fit values difficult. The drift in their values for $k_\text{max}\gtrsim 0.2\ihMpc$ may be interpreted as an indication that even higher-order (three-loop) contributions would start to become important there. This interpretation is also consistent with the estimate of the theoretical errors discussed above, and shown in Fig.\,\ref{fig:symmbased}. Nevertheless, within the region of validity of the two-loop approximation ($k_\text{max}\lesssim 0.2\ihMpc$), and for values of $k_\text{max}\gtrsim 0.1\ihMpc$ that are large enough to allow for their calibration, the best-fit values are reasonably well determined and consistent.

Note that the absolute values of the EFT parameters are sensitive to the renormalization scheme, in particular our choice to subtract the double-hard limit off the SPT two-loop piece in \eqref{eq:B2Lren}, while including the unsubtracted one-loop SPT contribution. Within this scheme, we find that the value obtained from the EFT two-loop fit for $\gamma_{1/2-\text{loop}}$ are close to the well-known $\cssq|_{P_{1L}}$ EFT parameter, as determined from the one-loop power spectrum, and indicated by the light gray line in the upper panels of Fig.\,\ref{fig:ctr}. While this agreement is not necessary from a theoretical point of view, it indicates that the dominant EFT correction shows a certain degree of universality, related to an approximately universal shift in the UV sensitive one-loop integral $\int^\Lambda d^3q P_{11}(q)/q^2$.

\begin{figure}
    \centering
    \includegraphics[width=0.49\textwidth]{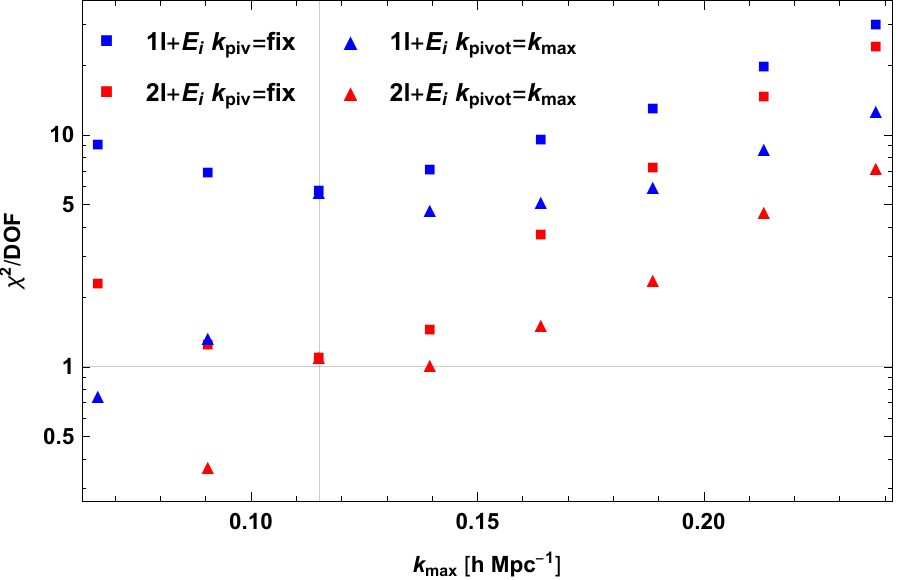}
    \caption{$\chi^2$/DOF as a function of $k_\text{max}$ for fixed (boxes, $k_\text{pivot}=0.115\ihMpc$) and running (triangles) parameter constraints.
    }
    \label{fig:chi2all}
\end{figure}

In order to further illustrate the impact of the two-loop terms, we show in Fig.\,\ref{fig:chi2all} the value of $\chi^2$/DOF as a function of $k_\text{max}$, but keeping the EFT parameters fixed for all values of $k_\text{max}$ (boxes). We use a pivot scale $k_\text{pivot}=0.115\ihMpc$ in that case, and also show the previous results with running EFT parameters in comparison (triangles). Clearly, both agree for $k_\text{max}=k_\text{pivot}$, and the $\chi^2$ with fixed parameters is necessarily worse as compared to running EFT parameters for $k_\text{max}\not =k_\text{pivot}$. For small $k_\text{max}$ the number of triangles is relatively small, such that the fit can account for possible fluctuations in the N-body data. In addition, the growth factor corrections, that affect also the tree-level bispectrum, tend to become more important there. This explains why the $\chi^2$ values with fixed parameters are significantly larger in that regime. For wavenumbers at which the two-loop correction gives a sizeable contribution, but higher orders are not yet important (around $0.1\ihMpc\lesssim k_\text{max}\lesssim 0.2\ihMpc$), the $\chi^2$ values with fixed or running EFT parameters are comparable, consistent with the finding that their running is small within this regime in the two-loop case. On the other hand, for the one-loop approximation the $\chi^2$ increases more strongly for $k_\text{max}\gtrsim k_\text{pivot}$ when using fixed EFT parameters. The difference between the $\chi^2$ with fixed and running EFT parameters (i.e.\ boxes versus triangles of the same color) for $k_\text{max}>k_\text{pivot}$ can therefore be interpreted as the tendency of the fit to absorb missing higher-order corrections into the EFT parameters. It is reassuring that this difference is small for the two-loop approximation (in red) within the relevant range of wavenumbers. In addition, one may argue that the difference in the $\chi^2$ between the one- and the two-loop approximation for fixed EFT parameters (i.e.\ blue versus red boxes) is a more faithful measure of the improvement when adding two-loop terms as compared to the running case (blue versus red triangles). 

Altogether, we find that the symmetry-based approach is able to describe the bispectrum with high precision up to $k_\text{max}\lesssim 0.08(0.15)\ihMpc$ in the EFT one-(two-)loop approximation, and that the relevant EFT parameters can be reliably calibrated using the grid-PT method within the respective regime of validity.

\begin{figure}
    \centering
    \includegraphics[width=0.49\textwidth]{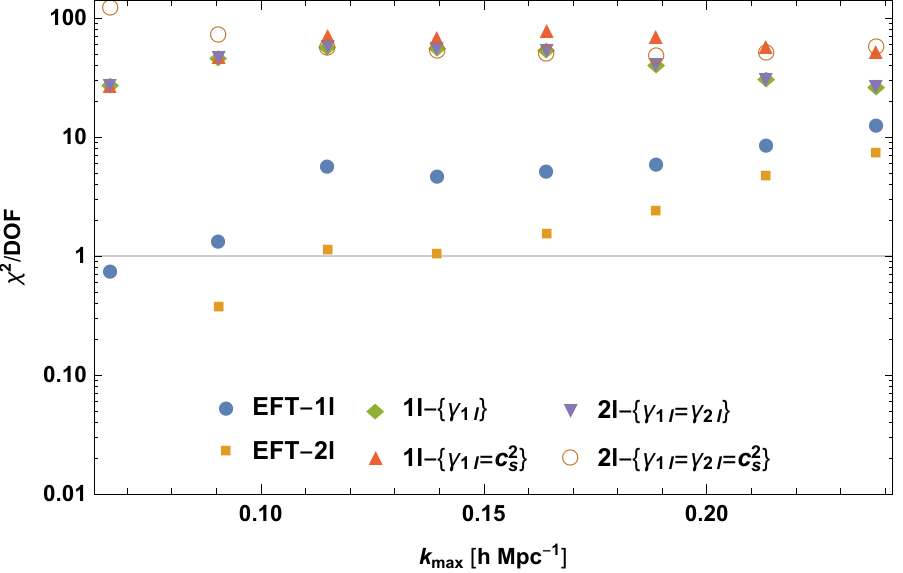}
    \caption{$\chi^2$/DOF for the UV-inspired one- and zero-parameter ansatz for the one- and two-loop bispectrum, respectively. Diamonds correspond to the one-parameter one-loop ansatz \eqref{eq:1LUV1par}, upward pointing triangles to the zero-parameter one-loop case \eqref{eq:1LUV0par}, downwards pointing triangles to the one-parameter two-loop ansatz \eqref{eq:2LUV1par}, and open circles to the zero-parameter two-loop ansatz \eqref{eq:2LUV0par}. The UV-inspired approach is included for illustration only, and does not account for the full UV dependence of the two-loop bispectrum. The proper symmetry-based EFT results discussed previously are also shown for comparison (filled circles and boxes).
    }
    \label{fig:chi2_uvinspired}
\end{figure}

\begin{figure}
    \centering
    \includegraphics[width=0.49\textwidth]{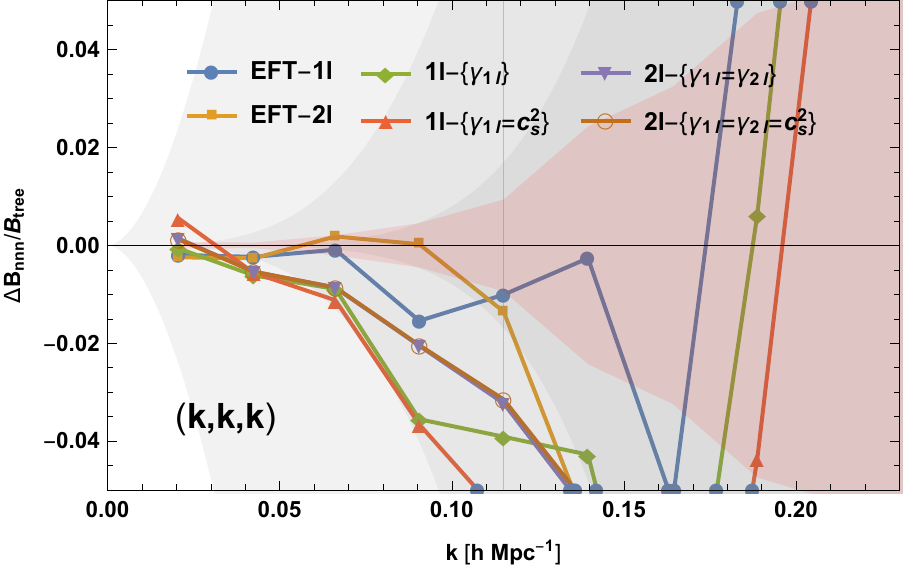}
    \includegraphics[width=0.49\textwidth]{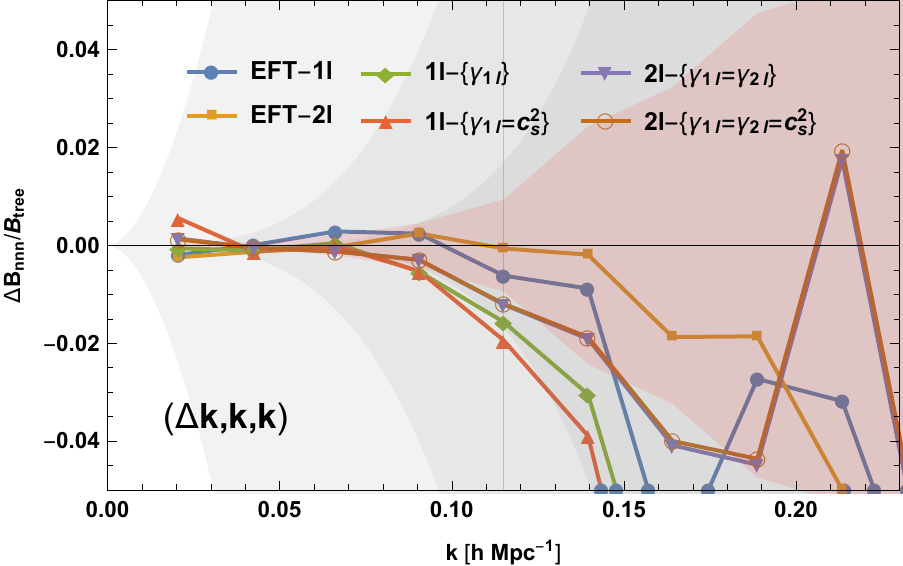}
    \includegraphics[width=0.49\textwidth]{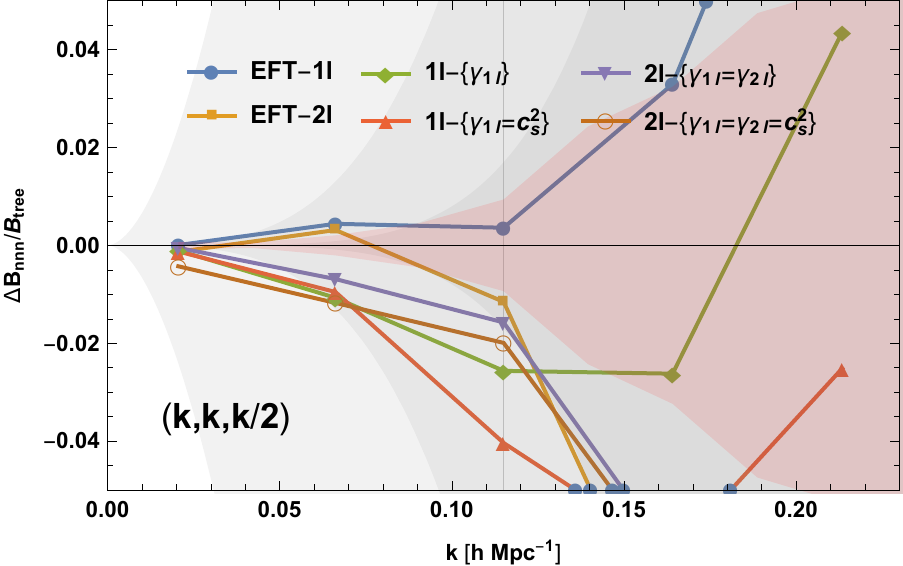}
    \includegraphics[width=0.49\textwidth]{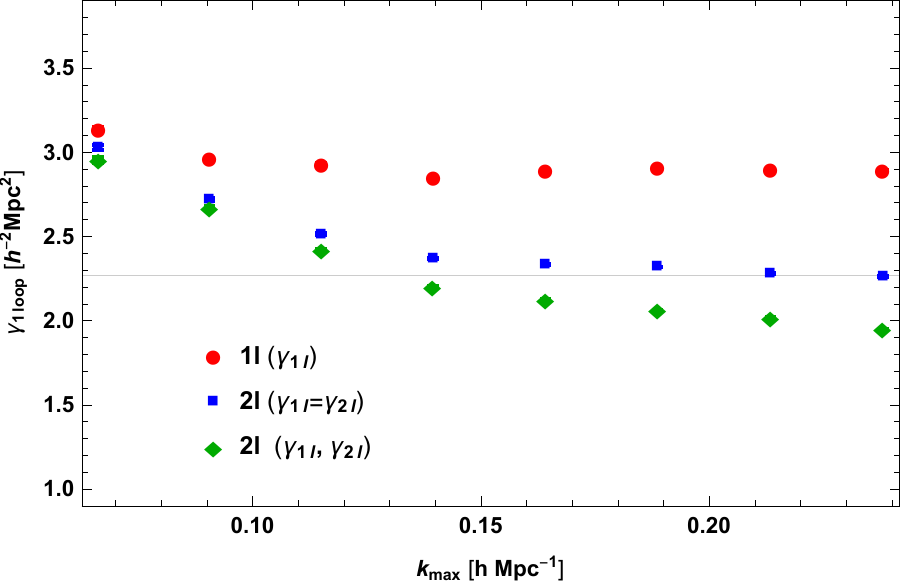}
    \caption{\emph{Upper row and lower left panel}: Relative deviation of the UV-inspired one- and zero-parameter ansatz for the one- and two-loop bispectrum, respectively, from N-body simulation results. Labels are as in  Fig.\,\ref{fig:chi2_uvinspired}, and shaded regions as in Fig.\,\ref{fig:symmbased}. The proper symmetry-based EFT results discussed previously are also shown for comparison. \emph{Lower right panel:} Best-fit value of $\gamma_{1-\text{loop}}$ for the one-parameter one- and two-loop UV-inspired approach, respectively. For illustration, the diamonds show $\gamma_{1-\text{loop}}$ for an extension to a two-parameter ansatz, with $\gamma_{1-\text{loop}}$ and $\gamma_{2-\text{loop}}$ treated as independent parameters.}
    \label{fig:uvinspired}
\end{figure}

\subsection{Comparison to approximate UV-inspired approach}

As discussed above, the rationale for using the UV-inspired approach (see \eqref{eq:1LUV1par}-\eqref{eq:2LUV0par}) is its simplicity, being described by either one or no extra EFT parameter compared to the one-loop power spectrum. Indeed, our analytical results for the double-hard limit of the bispectrum show that this reduction of free EFT parameters below the number of independent operators as dictated by the symmetry of the system is not appropriate. We compare the $\chi^2$ values for the one- and zero-parameter UV-inspired approach at one- and two-loop order, respectively, in Fig.\,\ref{fig:chi2_uvinspired}. It is apparent that $\chi^2$ is significantly larger for all UV-inspired cases as compared to the proper symmetry-based approach discussed previously, and shown again in Fig.\,\ref{fig:chi2_uvinspired} for comparison.

The relative deviation of the bispectrum from N-body data is shown in Fig.\,\ref{fig:uvinspired}, for the same shapes as in Fig\,\ref{fig:symmbased}. Apart from the symmetry-based approach (labeled EFT-1l/2l), we show the results obtained for four UV-inspired cases \eqref{eq:1LUV1par}-\eqref{eq:2LUV0par}. We find that the UV-inspired ansatz does not match the accuracy of the symmetry-based approach in all cases. In addition, the deviation starts already at relatively low wavenumbers, where a better agreement would be expected based on the estimated theoretical error, especially at two-loops (darkest gray shaded area). 

The lower right panel of Fig.\,\ref{fig:uvinspired} shows the best-fit values of the EFT parameter $\gamma_{1-\text{loop}}$ for the one-parameter UV-inspired one- and two-loop ansatz \eqref{eq:1LUV1par} and \eqref{eq:2LUV1par}, respectively, shown by the circles and squares. The EFT parameter shows a stronger running as compared to the proper symmetry-based approach, especially for the two-loop approximation (compare to upper left panel of Fig.\,\ref{fig:ctr}). For comparison, we also show the value of $\gamma_{1-\text{loop}}$ that is obtained in a generalization of the UV-inspired two-loop ansatz featuring two free parameters, i.e. allowing $\gamma_{1-\text{loop}}$ and $\gamma_{2-\text{loop}}$ to vary independently (diamonds). The value is consistent with the one obtained within the UV-inspired one-parameter ansatz \eqref{eq:2LUV1par} (squares) for the range of wavenumbers $k\lesssim 0.15 \ihMpc$ where the two-loop approximation is expected to work well. Consequently, an extension of the UV-inspired ansatz from one to two free parameters does not yield significant differences, and we therefore do not show further results for the latter.

In summary, we find that the simplified UV-inspired one- or zero-parameter ansatz for the bispectrum is not sufficient at two-loop order, while the symmetry-based approach discussed above yields a reliable EFT description of the bispectrum independent of the configuration. Nevertheless, we observe that the UV-inspired approach still yields a considerable improvement over SPT (see Fig.\,\ref{fig:symmbased}), while introducing a minimal set of free parameters, and could therefore be useful when comparing to data with a finite precision in practice.

\section{Conclusions}\label{sec:conclusions}

In this work we provide first results for the bispectrum at NNLO (two-loop) order in perturbation theory within the framework of an effective field theory description for the weakly non-linear modes in large-scale structure formation. We compute the two-loop bispectrum directly using a Monte Carlo integration scheme, as well as on the field level, known as grid-PT, allowing for a precise comparison to N-body simulations. Adding the two-loop correction increases the range of wavenumbers over which perturbation theory and N-body results agree at the percent level, and independently of the configuration of wavenumbers entering the bispectrum, from about $0.08\ihMpc$ at one-loop to $0.15\ihMpc$ at two-loop order. In particular, we find that the two-loop terms in the bispectrum start to become relevant already at $k\approx 0.07\ihMpc$.

The EFT description accounts for the UV sensitivity of standard perturbation theory, and we demonstrate independence of our results from the choice of the UV cutoff. At two-loops, one needs to consider the single- and double-hard limit of the loop integrals, with one or both integration variables becoming large, respectively. We provide analytical results for the double-hard limit, including in particular an expansion of the $F_6$ kernel entering $B_{611}$ for large loop wavenumbers. We show that the associated shape functions, that describe the dependence of the double-hard limit on the external wavenumbers $k_1,k_2,k_3$, can be mapped to the four second-order EFT operators known from the one-loop bispectrum. As may be expected in general, we find that the relative size of the coefficients of these four shapes are different for the hard limit of the one-loop bispectrum and the double-hard limit of the two-loop bispectrum, respectively. 

This implies that, within the EFT approach, the EFT parameters associated to the four second-order operators must be treated as independent from each other in order to be able to correct for the spurious UV sensitivity of SPT. 
In addition, we introduce one more parameter to account for the single-hard limit of the two-loop bispectrum. Our numerical results suggest that this parametrization is sufficient, while the most general EFT terms allowed by symmetries would correspond to the full set of up to fourth order EFT operators inserted into a one-loop bispectrum diagram. An extension of the EFT to this order goes beyond the scope of this work, and is left for future work.

Apart from the two-loop results, we also presented the first evaluation of the one-loop bispectrum using non-linear kernels evaluated for the precise $\Lambda$CDM cosmology, i.e.\ going beyond the EdS-SPT approximation for $F_4$. The corrections are comparable to the two-loop corrections, and therefore should be taken into account when working at NNLO level. 

Altogether, our results demonstrate that EFT methods allow for a systematic description of correlation functions order by order in perturbation theory, with controlled theoretical uncertainties. We find that the two-loop corrections are quantitatively relevant for wavenumbers that will be probed in future galaxy surveys, motivating an extension to the bispectrum of biased tracers.

\begin{acknowledgements}

We thank E. Pajer, R. Scoccimarro, M. Simonovic and Z. Vlah for helpful discussions and K. Kornet for excellent computing support. This research made use of the COSMOS supercomputer at the Department of Applied Mathematics and Theoretical Physics, Cambridge as well as the theory computing cluster at the Technical University of Munich supported by the ORIGINS excellence cluster. TB is supported by the Stephen Hawking Advanced Fellowship at the Center for Theoretical Cosmology. MG and PT are supported by the DFG Collaborative Research Institution Neutrinos and Dark
Matter in Astro- and Particle Physics (SFB 1258).
This project was initiated during a program hosted by the Munich Institute for Astro- and Particle Physics (MIAPP) which is funded by the
DFG – EXC-2094 – 390783311.

\end{acknowledgements}


%

\end{document}